\documentclass[aps,prd,reprint,amsmath,amssymb,nofootinbib,superscriptaddress,floatfix]{revtex4-2}
\usepackage{iftex}
\usepackage{mathrsfs}
\usepackage{graphicx}
\graphicspath{{image/}}
\usepackage[unicode,hidelinks]{hyperref}
\hypersetup{
  pdftitle={Single and
repetitive Penrose processes in genuine uncharged Kerr-Bertotti-Robinson black hole},
  pdfauthor={},
  pdfsubject={English manuscript draft},
  pdflang={en-US}
}

\makeatletter
\newcommand{\figurepanellabel}[2]{%
  \begingroup
  \protected@edef\@currentlabel{\thefigure(#2)}%
  \label{#1}%
  \endgroup
}
\makeatother
\begin{document}
\raggedbottom
% Suppress REVTeX's date prefix for the empty date.
\let\paperdate\date
\renewcommand{\date}[1]{\paperdate[]{#1}}
\title{Single and
repetitive Penrose processes in genuine uncharged\\ Kerr-Bertotti-Robinson black hole}
\author{Ruo-Cen Wu}
\email[]{wuruocen24@mails.ucas.ac.cn}
\affiliation{School of Fundamental Physics and Mathematical Sciences, Hangzhou Institute for Advanced Study, University of Chinese Academy of Sciences, Hangzhou 310024, China}
\affiliation{Taiji Laboratory for Gravitational Wave Universe (Beijing/Hangzhou), University of Chinese Academy of Sciences, Beijing 100049, China}
\affiliation{Institute of Theoretical Physics, Chinese Academy of Sciences, Beijing 100190, China}
\author{Hanjun Zou}
\email[]{zouhanjun24@mails.ucas.ac.cn}
\affiliation{School of Fundamental Physics and Mathematical Sciences, Hangzhou Institute for Advanced Study, University of Chinese Academy of Sciences, Hangzhou 310024, China}
\affiliation{Taiji Laboratory for Gravitational Wave Universe (Beijing/Hangzhou), University of Chinese Academy of Sciences, Beijing 100049, China}
\affiliation{Institute of Theoretical Physics, Chinese Academy of Sciences, Beijing 100190, China}
\author{Meng Xu}
\email[]{xumeng241@mails.ucas.ac.cn}
\affiliation{School of Fundamental Physics and Mathematical Sciences, Hangzhou Institute for Advanced Study, University of Chinese Academy of Sciences, Hangzhou 310024, China}
\affiliation{Taiji Laboratory for Gravitational Wave Universe (Beijing/Hangzhou), University of Chinese Academy of Sciences, Beijing 100049, China}
\affiliation{Institute of Theoretical Physics, Chinese Academy of Sciences, Beijing 100190, China}
% \author{Zhoujian Cao}
% \email[Zhoujian Cao:~]{zjcao@bnu.edu.cn}
% \affiliation{School of Physics and Astronomy, Beijing Normal University, Beijing 100875, China}
% \affiliation{School of Fundamental Physics and Mathematical Sciences, Hangzhou Institute for Advanced Study, University of Chinese Academy of Sciences, Hangzhou 310024, China}
\author{Yun-Long Zhang}
\email[]{zhangyunlong@nao.cas.cn}
\affiliation{National Astronomical Observatories, Chinese Academy of Sciences, Beijing 100101, China}
\affiliation{School of Fundamental Physics and Mathematical Sciences, Hangzhou Institute for Advanced Study, University of Chinese Academy of Sciences, Hangzhou 310024, China}
\date{}
\begin{abstract}
We establish an analytical framework for the Penrose process in the genuine uncharged Kerr-Bertotti-Robinson (KBR0) black-hole spacetime within a Bertotti-Robinson (BR) electromagnetic vacuum, consistently linking particle decay, radial transport, thermodynamic states, and geometric reconstruction by distinguishing coordinate Killing energy from thermodynamic mass-generator energy and performing multibranch parameter mapping from thermodynamic states to background metric parameters.
Furthermore, we present the analytical solution for the decay of equatorial massive neutral particles at a common radial turning point inside the ergoregion, the orbital accessibility criteria, and the six physical constraints required for a complete extraction event.
Numerical calculations demonstrate that for both near-extremal and lower-spin inputs, in both single and repetitive Penrose processes, the net energy return on investment ($\eta_{\text{EROI}}$) and the rotational energy utilization efficiency ($\eta_{\text{EUE}}$) decrease with increasing external magnetic field, whereas the rotational energy reservoir consumption rate per unit incident energy ($\eta_{\text{thermo}}$) increases.
For the inputs considered, the feasible domain and the upper bound of the initial magnetic field for single extraction are constrained by two boundary mechanisms: the obstruction of parent-particle infall and the loss of negative energy in the captured fragment.
Finally, for repetitive extraction processes within the surveyed regime, all sequences terminate due to orbital accessibility obstructions, explicitly local-slope failure and far-field incident-branch closure, with the final black-hole state retaining most of its initial rotational energy reservoir. Moreover, the black-hole state evolves along a common integral curve independent of the parent particle's specific energy, which solely alters the sequence endpoint.
\end{abstract}
\maketitle

\tableofcontents
\section{Introduction}
\label{sec:introduction}

Extracting rotational energy from rotating black holes is a fundamental problem in energy exchange in strong gravitational fields and in the mechanisms powering high-energy astrophysical phenomena. The Penrose process identifies a physical route to extracting this energy through negative-energy states inside the ergoregion~\cite{Penrose:1969pc,Penrose:1971uk}: when the black hole captures fluxes carrying negative energy and counterrotating angular momentum, external matter or fields can gain a net energy output. Both purely gravitational mechanical decay of particles and energy-extraction processes involving background electromagnetic fields~\cite{Bhat:1985,Wagh:1989zqa,Tursunov:2019,Zeng:2025reconnection} can be described within this common conserved-flux framework~\cite{lasota_generalized_2014}. However, the theoretical upper limit on extractable rotational energy~\cite{PhysRevLett.25.1596,PhysRevD.4.3552}, the net energy gain from a single decay~\cite{Wald:1974limits}, and the cumulative net output of an evolving extraction sequence~\cite{ruffini_single_2025,ruffini_role_2025} are physically distinct quantities. In repetitive extraction, in particular, the black hole's thermodynamic mass, angular momentum, and irreducible mass continually change through capture, while the orbital accessibility conditions for external particles are updated dynamically. Studies of single and repetitive Penrose processes, including their electric generalizations, in asymptotically flat black holes~\cite{ruffini_single_2025,ruffini_role_2025,Alipour:2026kn,Kumar:2026sen,Li:2026gmghs} and generalizations to other black-hole geometries~\cite{Wang:2025nfv,Zeng:2026egq,Zeng:2026jpn,Alloqulov:2026cyb,Alipour:2026egb,Sabir:2026rastall,Afshar:2026sv}, together with related analyses~\cite{Rueda:2024xeb,hu_third_law_2026,Feiteira:2026recursive,Wang:2026jah}, highlight the importance of irreversible growth of the irreducible mass and feedback from evolving black-hole parameters and electromagnetic fields in assessing energy-recovery efficiency.

Extending this energy-extraction picture to magnetized black-hole spacetimes that are not asymptotically flat first requires clarifying how the external magnetic field affects particle motion through the background geometry~\cite{Wang:2025geodesics,Wang:2025isco} and how the thermodynamic quantities are defined~\cite{kubiznak_thermodynamics_2026,zhang_thermodynamics_2026}. The exact Kerr--\allowbreak Newman--\allowbreak Bertotti--\allowbreak Robinson (KNBR) family provides a rigorous analytical framework for this purpose. Podolsk\'{y} and Ovcharenko first constructed the Kerr--BR solution describing a rotating black hole immersed in a uniform Bertotti--\allowbreak Robinson (BR) magnetic field~\cite{podolsky_kerr_2025}. The subsequent general KNBR construction fully decoupled the black-hole charge from the external-field parameter, thereby clearly distinguishing the special charged branch represented by the earlier solution from the genuine uncharged branch with strictly vanishing physical charge~\cite{ovcharenko_kerr-newman_2026}. The essential difference between KNBR and standard Kerr--Newman (KN) spacetimes originates in the change of background vacuum: the standard KN solution asymptotically approaches the flat Minkowski vacuum, whereas the KNBR solution is embedded in the non-asymptotically flat BR electromagnetic vacuum~\cite{Bertotti:1959,Robinson:1959,ovcharenko_kerr-newman_2026}. Since the background electromagnetic field and its gravitational backreaction directly form part of the spacetime geometry, the external magnetic field reshapes the event horizon and ergoregion near the black hole~\cite{podolsky_kerr_2025,ovcharenko_kerr-newman_2026}. Its effects also extend to the exterior geometry and global structure~\cite{Zhou:2026tkm,Ovcharenko:2026ooh,Zhou:2026fvj,Zhou:2026extended}, which enter the interpretation of particle incidence, energy output, and the normalization of conserved quantities~\cite{ovcharenko_kerr-newman_2026,kubiznak_thermodynamics_2026}.

The study of demagnetizing transformations by Ma and L\"u further clarifies the intrinsic relation between geometric deformation and electromagnetic parameters in these spacetimes~\cite{ma_demagnetizing_2026}. By applying a demagnetizing transformation to the earlier KBR solution, they obtained a Ricci-flat rotating seed metric that retains a geometric deformation parameter after the electromagnetic field is removed and has no asymptotically flat region. Upon remagnetization, the external magnetic-field strength can be varied independently of this geometric deformation parameter. This demagnetization construction provides a richer comparison than directly taking the $B=0$ limit of the original metric. It also shows that in energy-extraction studies in magnetized backgrounds, the relationships among external-field strength, background geometric deformation, and physical parameter identification must be carefully delineated. The demagnetized Ricci-flat seed metric and the genuine uncharged branch studied here, which remains in a BR electromagnetic vacuum, represent different physical settings and must be treated separately.

Research on KNBR spacetimes has now expanded from exact geometric constructions to black-hole thermodynamics and particle energy extraction. The thermodynamic properties of the original, specially charged Kerr--BR branch have been studied~\cite{Hu:2026slp,Hassanabadi:2026fzz}. In particular, work on genuine uncharged Kerr--BR black holes introduces an appropriately normalized time generator and establishes a consistent thermodynamic mass expression, first law, and Smarr relation~\cite{kubiznak_thermodynamics_2026}. Within the covariant phase-space framework~\cite{Iyer:1994ys}, work on general charged KNBR black holes determines the thermodynamic mass and potentials for independent charge and external-field parameters~\cite{zhang_thermodynamics_2026}. A subsequent analysis derives the same mass from charge integrability without assuming thermodynamic relations and independently verifies the first law and Smarr formula~\cite{Hu:2026integrable}. Meanwhile, the magnetic Penrose process and charged-particle acceleration~\cite{mirkhaydarov_mpp_2026}, as well as energy extraction through the capture of negative-energy particles~\cite{wang_energy_2026}, have also been discussed on related Kerr--BR branches. These developments make the consistent combination of background geometry, thermodynamic evolution, and global particle transport a key problem in further studies of energy extraction from KNBR black holes.

Against this background, we focus on the neutral-particle Penrose process on the genuine uncharged KNBR branch, hereafter denoted KBR0. Choosing neutral test particles means that the external magnetic field affects their geodesic motion only through the background geometry, allowing us to isolate and examine how the magnetized spacetime geometry itself controls energy extraction. We therefore address three central questions. First, how can local negative-energy decay inside the ergoregion be connected globally to a prescribed outer boundary and the future event horizon? Second, how can the normalization difference between coordinate Killing energy and the thermodynamic mass generator be accounted for so that the captured fluxes and black-hole thermodynamic quantities use a consistent energy generator? Third, as the background evolves step by step during extraction, how can single-event net gain, irreversible consumption of rotational energy, and cumulative sequence output be compared quantitatively and consistently? Therefore we must also distinguish comparisons at fixed initial thermodynamic state from those at fixed initial metric parameters and fully retain the geometric branch information when reconstructing metric parameters from each updated thermodynamic state.

To this end, we consider instantaneous decay of equatorial massive neutral test particles at a common radial turning point, with the external magnetic field held fixed, and organize a single extraction step as the state-update chain:
\begin{equation}
(m,a)\longrightarrow(M,J)\longrightarrow(M',J')
\longrightarrow(m',a').
\label{eq:process-chain}
\end{equation}
We first compute the black-hole thermodynamic state from the current metric parameters, then update its thermodynamic mass and angular momentum using the conserved fluxes carried by the captured fragment, and finally recover the updated background geometry through parameter mapping from the thermodynamic state on the selected algebraic branch. At each extraction step, we jointly check six physical criteria: local mass-shell conditions and four-momentum conservation, the future time orientation of all three particles, negative energy of the captured fragment, global incidence of the parent and transport accessibility of both fragments on the background, and the validity of the updated background geometry together with nondecreasing outer-horizon area. A complete extraction event satisfying all these criteria is defined as a single Penrose process (SPP). A sequence that successively updates the black-hole background~\cite{ruffini_role_2025} and continues until a physical obstruction is first triggered is defined as a repetitive Penrose process (RPP). Its terminal state is the background after the last completed update before the obstruction.

Specifically, we first establish the analytical relations for parameter mapping and branch selection from the KBR0 thermodynamic state to background metric parameters, derive the analytical particle decay solution and global radial transport criteria for the common-turning-point configuration, and define efficiency measures for single events and repetitive sequences that account for both net external output and consumption of the rotational energy reservoir. In the numerical analysis, we first scan the external magnetic-field strength at two fixed initial thermodynamic states, one near extremality and the other at lower spin, systematically comparing extraction sequence lengths, energy partition, and orbital termination mechanisms across field strengths. We then vary the parent particle's incident specific energy at fixed external magnetic field to examine independently the relation between the black-hole state evolution path and the attainable sequence endpoint. All these comparisons use consistent particle-input parameters, outer-boundary settings, and branch-selection conventions for parameter mapping.

The remainder of this paper is organized as follows. Section~\ref{sec:geometry} introduces the KBR0 background geometry, the definitions of thermodynamic quantities, and the multibranch algebraic reconstruction of background metric parameters. Section~\ref{sec:process} presents the six physical conditions for a complete extraction event, the particle decay solution, and the radial transport criteria. Section~\ref{sec:efficiency} defines the single-event and cumulative extraction efficiencies. Section~\ref{sec:results} presents the numerical results for single and repetitive extraction and analyzes the physical mechanisms. Section~\ref{sec:conclusion} summarizes the conclusions and discusses possible theoretical extensions. The appendices provide detailed mathematical derivations of parameter mapping and its geometric branches, the particle decay solution, fragment radial transport, restrictions on the decay-parameter window, and the symmetry under spin reversal and axial reflection.
\section{Geometry and thermodynamics}
\label{sec:geometry}

The KBR0 background is characterized by the geometric state $(m,a,B)$, with line element~\cite{ovcharenko_kerr-newman_2026}
\begin{equation}
\begin{aligned}
\mathrm{d}s^2=\frac{1}{\mathcal W_{\text{c}}^2}\Bigl[&-\frac{Q}{\rho^2}
(\mathrm{d}t-a\sin^2\theta\,\mathrm{d}\varphi)^2
+\frac{\rho^2}{Q}\mathrm{d}r^2\\
&+\frac{\rho^2}{P}\mathrm{d}\theta^2
+\frac{P\sin^2\theta}{\rho^2}
(a\,\mathrm{d}t-(r^2+a^2)\,\mathrm{d}\varphi)^2\Bigr],
\end{aligned}
\label{eq:metric}
\end{equation}
where the remaining functions are given in Appendix~\ref{app:geometry}. Removing the conical singularities on the axis fixes the period of $\varphi$ to $2\pi C$. We use the regular azimuthal angle $\phi=K\varphi$, with $C=K^{-1}$, so that $\phi$ has period $2\pi$. Useful background quantities are
\begin{equation}
\begin{gathered}
b^2=1+m^2B^2,\qquad \mu_{\text{h}}^2=m^2-a^2b^4\ge0,\\
K=1+B^2\mu_{\text{h}}^2,\qquad r_\pm=(m\pm\mu_{\text{h}})/b^2.
\end{gathered}
\label{eq:background}
\end{equation}
We take $m>0$, $b>0$, and $\mu_{\text{h}}\ge0$, allowing both spin orientations. The correspondence between spin reversal and axial reflection is given in Appendix~\ref{app:spin-reflection}. The equatorial ergoregion satisfies $r_+<r<r_{\text{ergo}}$, where its outer boundary is the unique exterior root of $Q(r_{\text{ergo}})=a^2$, and no ergoregion exists for $a=0$.

The normalization of the mass generator is fixed by integrability on the solution space and the Kerr limit~\cite{kubiznak_thermodynamics_2026,zhang_thermodynamics_2026,Hu:2026integrable}. On the uncharged branch considered here,
\begin{equation}
\begin{gathered}
\xi_M=b^{-1}\partial_t,\qquad
M=\frac{mC}{b},\qquad J=maC^2,\\
S=\frac{\pi C(r_+^2+a^2)}{I_+},\qquad
T=\frac{\mu_{\text{h}} I_+}{2\pi b(r_+^2+a^2)},\\
\Omega=\frac{a}{Cb(r_+^2+a^2)},\qquad I_+=I(r_+).
\end{gathered}
\label{eq:thermodynamic-state}
\end{equation}
These quantities satisfy the Christodoulou--Ruffini relation~\cite{PhysRevLett.25.1596,PhysRevD.4.3552}, the first law~\cite{Bardeen:1973laws}, and the Smarr relation~\cite{Smarr:1973mass,kubiznak_thermodynamics_2026},
\begin{equation}
\begin{aligned}
M^2&=\frac{S}{4\pi}+\frac{\pi J^2}{S},\\
\mathrm{d}M&=T\,\mathrm{d}S+\Omega\,\mathrm{d}J,\\
M&=2TS+2\Omega J.
\end{aligned}
\label{eq:thermodynamic-relations}
\end{equation}
The outer-horizon area selects $S=2\pi[M^2+\sqrt{M^4-J^2}]$. This uniquely determines the irreducible mass $M_{\text{irr}}=\sqrt{S/(4\pi)}$, while the branch information of the background geometry is retained during parameter mapping as defined below.

For a given thermodynamic state $(M,J,B)$, the parameter mapping determines the metric parameters on a selected geometric branch. We refer to this recovery of metric parameters from the thermodynamic state as parameter mapping. Introduce the dimensionless variables
\begin{equation}
j=J/M^2,\quad\delta=1-j^2,\quad
\beta=(BM)^2,\quad x=m^2/M^2.
\label{eq:parameter-mapping-variables}
\end{equation}
The parameter $\delta$ measures departure from extremal spin. Equations~\eqref{eq:background} and~\eqref{eq:thermodynamic-state} give $\delta=\mu_{\text{h}}^2/m^2$, hence $0\le\delta\le1$ on the positive-mass branch with a horizon. The case $\delta=0$ denotes an extremal state in which the inner and outer horizons coincide, whereas $\delta=1$ denotes the nonrotating state $J=a=0$. Smaller $\delta$ corresponds to a smaller normalized horizon separation, $(r_+-r_-)/(r_++r_-)=\sqrt{\delta}$. This parameter links thermodynamic spin to geometric horizon degeneracy, but does not distinguish different geometric root branches at the same $(M,J,B)$. The mass and angular-momentum equations reduce to
\begin{equation}
\mathcal F_{\delta,\beta}(x)
=x-(1+\beta x)(1+\delta\beta x)^2=0,\quad x>0.
\label{eq:parameter-mapping-cubic}
\end{equation}
Each positive root gives $m=M\sqrt{x}$ and $a=jM\sqrt{x}/(1+\beta x)$. For $0<\delta\le1$ and $\beta>0$, the fold at which the positive root becomes double is
\begin{equation}
u_*=\frac{\sqrt{1+8/\delta}-1}{4},\qquad
\beta_* =\frac{u_*}{(1+u_*)(1+\delta u_*)^2}.
\label{eq:fold}
\end{equation}
The cases $\beta<\beta_*$, $\beta=\beta_*$, and $\beta>\beta_*$ respectively correspond to two distinct positive roots, one double positive root, and no positive root. The two geometric roots share the same outer-horizon entropy, but each must still pass the exterior-geometry checks. Quasistatic small steps at fixed $B$ use continuous continuation on the same branch and stop at the fold or when the tracked root becomes invalid. Our numerical calculations start on the smaller-root branch continuously connected to Kerr. Elimination, degenerate boundaries, and the basis for branch selection are detailed in Appendix~\ref{app:parameter-mapping}.

Table~\ref{tab:notation} summarizes the notation used below. Three mass scales must be distinguished: $m$ is the metric parameter, $M$ is the current thermodynamic mass, and $M_0$ is the fixed initial reference mass of a sequence. The quantity $\beta=(BM)^2$ evolves with the background, whereas $\beta_0=(BM_0)^2$ labels the initial magnetic field. An iterative Penrose process at fixed $B$ therefore still changes $\beta$. These conventions are used throughout the particle dynamics and efficiency comparisons.

\begin{table*}[tbp]
\caption{Core notation and normalization conventions. The subscript $0$ denotes the initial state for black-hole quantities. For particle quantities, the labels $0,1,2$ denote the parent, captured fragment, and outgoing fragment, respectively.}
\label{tab:notation}
\begin{ruledtabular}
\begin{tabular}{p{0.15\textwidth}p{0.15\textwidth}p{0.65\textwidth}}
Category & Symbol & Meaning or definition \\
\hline
\raggedright Geometry and magnetic field & $m,a,B$ & Geometric mass parameter, rotation parameter, external magnetic field \\
 & $b,\mu_{\text{h}}$ & $b^2=1+m^2B^2$, $\mu_{\text{h}}^2=m^2-a^2b^4$(the horizon discriminant parameter) \\
 & $K,C,\mathcal W_{\text{c}}$ & Axial factors $K=1+B^2\mu_{\text{h}}^2$, $C=K^{-1}$, conformal factor $\mathcal W_{\text{c}}$ \\
 \hline
\raggedright Thermodynamics & $M,J,S,T,\Omega$ & Mass, angular momentum, outer-horizon entropy, temperature, thermodynamic angular velocity \\
 & $M_{\text{irr}},E_{\text{rot}}$ & Irreducible mass $\sqrt{S/(4\pi)}$, rotational energy reservoir $M-M_{\text{irr}}$ \\
 & $\xi_M,M_0$ & Mass generator $b^{-1}\partial_t$, fixed initial reference mass \\
 \hline
\raggedright Particles and local quantities & $\mu_{(i)},E_i,L_i$ & Particle rest mass, coordinate Killing energy, angular momentum for the regular axial coordinate \\
 & $\varepsilon_i,\ell_i$ & $\varepsilon_i=E_i/\mu_{(i)}$, $\ell_i=L_i/(m\mu_{(i)})$ \\
 & $E_{M,i},\bar\ell_1$ & Mass-generator energy $E_i/b$, $L_1/(M\mu_{(1)})$, with $M=M_0$ at the first step \\
 & $\chi_i,\Gamma_{01},\lambda,\sigma$ & ZAMO rapidity, relative Lorentz factor, $\mu_{(2)}/\mu_{(1)}$, $\mu_{(1)}/\mu_{(0)}$ \\
 \hline
\raggedright Dimensionless quantities & $j,\delta,\beta,x$ & $J/M^2$, $1-j^2$, $(BM)^2$, $m^2/M^2$ \\
 & $u,\beta_0,f$ & $B^2m^2=\beta x$, $(BM_0)^2$, $\beta_0/\beta_{0,\text{max}}$ \\
 \hline
\raggedright Orbits and sequences & $N(r),\omega(r),R_{\text{circ}}(r)$ & ZAMO lapse, local frame-dragging angular velocity, circumferential radius \\
 & $\Gamma_\infty,A_4,S_{\text{d}}$ & Exterior-geometry condition, leading radial coefficient at large radius, local radial slope at the turning point \\
 & $N$ & Number of completed extraction events, distinguished from $N(r)$ by context \\
\end{tabular}
\end{ruledtabular}
\end{table*}
\section{Kinematics and global extraction criteria}
\label{sec:process}

The previous section established the correspondence between background geometry and thermodynamic state and explained why branch information must be retained after an update. We now incorporate local particle decay and global radial motion into Eq.~\eqref{eq:process-chain}. A complete extraction event must satisfy all six conditions below:
\begin{enumerate}
\item The background has an outer event horizon, and the decay point lies in the ergoregion of a valid exterior geometry.
\item The parent particle can reach the decay turning point from the prescribed outer boundary.
\item All three particles satisfy their mass-shell conditions and local four-momentum conservation~\cite{Zhang:2026decay,Chen:2026decaykn}.
\item All three particles are future directed, and the captured fragment satisfies $E_1<0$.
\item Fragment 1 crosses the future horizon, and fragment 2 is connected to the prescribed output boundary.
\item The updated state can be reconstructed as a valid geometry on the selected branch, with nondecreasing outer-horizon area~\cite{Hawking:1971area}.
\end{enumerate}
We discuss the local solution, update, and radial transport in turn. In an RPP, all conditions are reevaluated after every completed extraction. The terminal state is the background after the last completed update before the first physical obstruction. In all SPP and RPP calculations performed here, the prescribed outer boundary is coordinate infinity.

\subsection{Particle decay and thermodynamical evolution}
\label{sec:local}

Parent particle 0 moves inward from the outer boundary, reaches a radial turning point $r_{\text{d}}$ inside the ergoregion, and splits there into fragments 1 and 2 with positive rest masses. We choose a common radial turning point~\cite{ruffini_single_2025,ruffini_role_2025}: at the instant of decay, all three particles have $p^r_{(i)}=0$, and their spatial momenta measured by an equatorial zero-angular-momentum observer (ZAMO)~\cite{1972ApJ...178..347B} are purely azimuthal. When the transport conditions below are satisfied, the incident parent orbit reaches its minimum radius at $r_{\text{d}}$, and fragment 1 then falls inward through the horizon, while fragment 2 moves outward. The particles share a turning position but have different subsequent radial directions. Define
\begin{equation}
E_i=-p_{(i)t},\quad L_i=p_{(i)\phi},\quad
\varepsilon_i=\frac{E_i}{\mu_{(i)}},\quad
\ell_i=\frac{L_i}{m\mu_{(i)}}.
\label{eq:particle-notation}
\end{equation}
The angular momentum is associated with the regular axial generator $\partial_\phi$, and $m$ is the geometric mass parameter of the background before decay at the current step.

For an equatorial ZAMO, the lapse, frame-dragging angular velocity, and circumferential radius are $N(r)$, $\omega(r)$, and $R_{\text{circ}}(r)$, respectively. Their explicit expressions are given in Appendix~\ref{app:geometry}. We denote their values at the decay radius by $N_{\text{d}}=N(r_{\text{d}})$, $\omega_{\text{d}}=\omega(r_{\text{d}})$, and $R_{\text{circ,d}}=R_{\text{circ}}(r_{\text{d}})$. The future mass shell at the decay point can be written in terms of signed rapidities as
\begin{equation}
\begin{gathered}
\varepsilon_i=N_{\text{d}}\cosh\chi_i+\omega_{\text{d}} R_{\text{circ,d}}\sinh\chi_i,
\\ m\ell_i=R_{\text{circ,d}}\sinh\chi_i.
\end{gathered}
\label{eq:rapidity}
\end{equation}
Once the background and the decay position $r_{\text{d}}$ inside the ergoregion are specified, $N_{\text{d}},\omega_{\text{d}},R_{\text{circ,d}}$ are determined. The parent specific energy $\varepsilon_0>0$ then uniquely determines its real future-directed rapidity $\chi_0$ through Eq.~\eqref{eq:rapidity}, while the prescribed $\ell_1$ uniquely determines $\chi_1$. Specifying the fragment mass ratio $\lambda=\mu_{(2)}/\mu_{(1)}>0$ allows four-momentum conservation and the future mass-shell condition of fragment 2 to determine the mass fraction $\sigma$. The parent rest mass $\mu_{(0)}$ is an independently prescribed overall scale. Once this scale is chosen, the fragment rest masses are fixed:
\begin{equation}
\begin{gathered}
\Gamma_{01}=\cosh(\chi_0-\chi_1),\qquad
\sigma=\frac{1}{\Gamma_{01}+\sqrt{\Gamma_{01}^2+\lambda^2-1}},\\
\mu_{(1)}=\sigma\mu_{(0)},\qquad
\mu_{(2)}=\lambda\sigma\mu_{(0)}.
\end{gathered}
\label{eq:mass-ratio}
\end{equation}
The condition $E_1<0$ requires the angular momentum of fragment 1 to oppose the black-hole spin. Fragment 2 is determined by $p_{(2)}=p_{(0)}-p_{(1)}$. The explicit solution and the selection of the future branch are given in Appendix~\ref{app:local}.

The captured quantities must use the same generators as the black-hole thermodynamic quantities. At a given step, the mass-generator energy is $E_{M,i}=-p_{(i)}(\xi_M)=E_i/b$, giving the small-step update
\begin{equation}
M'=M+E_1/b,\qquad J'=J+L_1,\qquad B'=B.
\label{eq:ledger}
\end{equation}
Here $b$ is evaluated on the background before decay. The final geometry is obtained by parameter mapping on the same branch using Eq.~\eqref{eq:parameter-mapping-cubic}, and the finite entropy difference $\Delta S\ge0$ is checked using the initial and final outer-horizon areas. This connects the local energy gain to the overall change of the black hole. We use a test-particle capture model at fixed $B$ and check the area condition separately as a finite difference. Self-force and radiation reaction are left to an extended model.

\subsection{Orbital accessibility and fragment transport}
\label{sec:transport}

After obtaining the local solution, the turning point must still be connected to an exterior incident orbit. Holding the conserved quantities $(\varepsilon,\ell)$ fixed along each geodesic, radial motion is governed by
\begin{equation}
\begin{aligned}
(u^r)^2&=I^2\hat{\mathcal R}/r^4,\\
\hat{\mathcal R}(r)&=[(r^2+a^2)\varepsilon-aKm\ell]^2\\
&\quad-Q(a\varepsilon-Km\ell)^2-\Delta r^2.
\end{aligned}
\label{eq:radial}
\end{equation}
The common turning point ensures $\hat{\mathcal R}(r_{\text{d}})=0$. Thus factoring out this root leaves a radial quotient of degree at most three. The condition for the valid geometry to cover the entire exterior radial half-line is $\Gamma_\infty=1-a^2B^2b^4\ge0$. Within this domain and on the selected future branch, the conditions for the parent to reach a simple turning point from coordinate infinity reduce to
\begin{equation}
A_4\ge0,\qquad S_{\text{d}}>0.
\label{eq:parent-access}
\end{equation}
Here $M_0$ is the fixed initial thermodynamic mass, $A_4$ is the quartic coefficient of the dimensionless radial function $\overline{\mathcal R}=\hat{\mathcal R}/M_0^4$ in $\varrho=r/M_0$, and $S_{\text{d}}=\partial_\varrho\overline{\mathcal R}|_{r_{\text{d}}/M_0}$. The derivative is taken at fixed orbital conserved quantities. These conditions check the large-radius and local behavior separately: $A_4\ge0$ keeps radial motion allowed at infinity, while $S_{\text{d}}>0$ makes the region immediately outside the turning point accessible. Both are required to deliver the parent to the decay point.

The condition $A_4<0$ indicates a radial barrier at large radius, while $S_{\text{d}}<0$ indicates a forbidden region immediately outside the decay point. The two equality boundaries have different meanings. At $A_4=0$, the radial polynomial reduces to a cubic with positive leading coefficient and can still support incidence. At $S_{\text{d}}=0$, the turning root becomes degenerate and the proper time required to reach it diverges. The full criterion is given in Appendix~\ref{app:radial}.

Stronger analytical results control the fragments. For a future-directed negative-energy turning fragment, the radial function strictly decreases with increasing radius on the inner interval. Motion inward from the turning point remains allowed and satisfies the strict future-horizon crossing condition. The velocity ordering implied by local conservation also ensures that the outward allowed region of the positive-energy fragment contains the parent's accessible region connected to the same boundary. Appendix~\ref{app:transport} proves these two properties. Thus, for neutral equatorial particles sharing a turning point and the same incident and output boundary, parent accessibility and local negative-energy decay jointly guarantee transport on the background. The finite final black-hole state remains subject to the sixth condition.

\subsection{Feasible extraction parameter domain}
\label{sec:windows}

\begin{figure*}[!t]
\centering
\includegraphics[width=\textwidth]{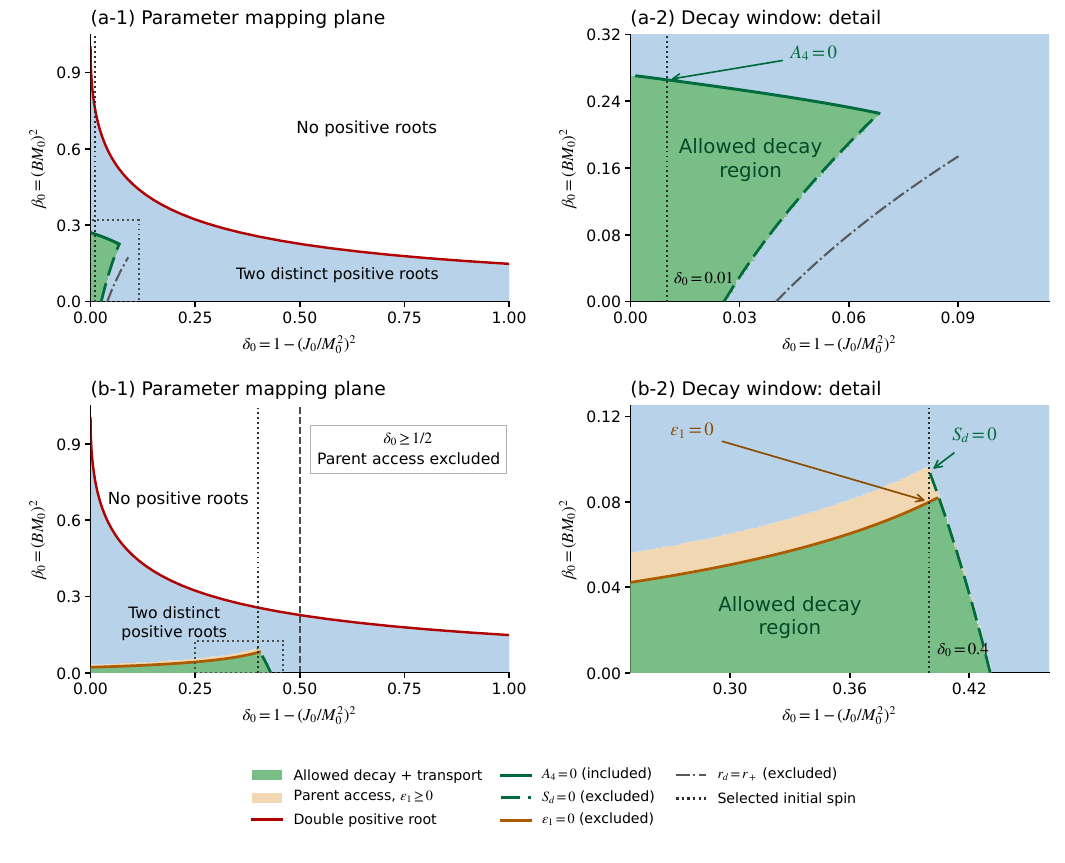}
\caption{Particle-accessible windows within the algebraic parameter mapping plane for $\varepsilon_0=2$, $\bar\ell_1=-10$, and $\lambda=\mu_{(2)}/\mu_{(1)}=0.6$. Panels (a-1) and (a-2) show the full plane and its enlargement for $r_{\text{d}}/M_0=1.2$; (b-1) and (b-2) do the same for $r_{\text{d}}/M_0=1.95$. The color and boundary key is below all four panels. White denotes the absence of a positive root of the parameter mapping equation. The axes use $\delta_0=1-(J_0/M_0^2)^2$ and $\beta_0=(BM_0)^2$. Green denotes local negative-energy decay with background transport; finite black-hole final states require a separate check.}
\label{fig:window}
\figurepanellabel{fig:window:a-1}{a-1}
\figurepanellabel{fig:window:a-2}{a-2}
\figurepanellabel{fig:window:b-1}{b-1}
\figurepanellabel{fig:window:b-2}{b-2}
\end{figure*}

Figure~\ref{fig:window} overlays the particle conditions for two decay positions on the parameter mapping plane. The two-positive-root region supplies geometric candidates, while parent incidence and fragment negative energy select a smaller domain. For $\varepsilon_{0} = 2, \bar{\ell}_{1} = -10, \lambda = \mu_{(2)}/\mu_{(1)} = 0.6$, the large-radius condition and local slope provide the main particle boundaries at $r_{\text{d}}/M_0=1.2$. At $r_{\text{d}}/M_0=1.95$, the finite counterrotating angular momentum can instead cause the fragment to lose negative energy while the parent remains accessible. Section~\ref{sec:results} uses $\delta_0=0.01$ and $0.4$ as numerical slices for these two cases, respectively.

The radial window yields two concise necessary restrictions. At fixed parent specific energy, the large-radius condition requires $B^2m^2\le\varepsilon_0^2-1$, so incidence from infinity at nonzero field uses $\varepsilon_0>1$. A common turning point inside the ergoregion also requires $|j_0|>1/\sqrt2$, hence $\delta_{0} < 1/2$. Once these necessary conditions hold, complete events are selected using Eq.~\eqref{eq:parent-access}, the negative-energy condition, and the final-state requirements. The geometric window, strict boundaries, and necessary lower spin bound are derived in Appendix~\ref{app:window}.

\section{Efficiency indices of energy extraction}
\label{sec:efficiency}

A decay event is included in the energy budget only after it passes the six conditions of the previous section. Its net energy return on investment (EROI) is defined as~\cite{hu_third_law_2026}
\begin{equation}
\eta_{\text{EROI}}=\frac{E_2-E_0}{E_0}=-\frac{E_1}{E_0}=-\frac{E_1/b}{E_0/b}.
\label{eq:eta-eroi}
\end{equation}
This measures net gain, whereas $E_2/E_0$ is the gross output-to-input ratio. To distinguish net gain from irreversible consumption, define $E_{\text{rot}}=M-M_{\text{irr}}$ and evaluate the finite difference between initial and final states:
\begin{equation}
-\Delta E_{\text{rot}}=\frac{|E_1|}{b}+\Delta M_{\text{irr}}.
\label{eq:reservoir-split}
\end{equation}
Part of the decrease in the rotational reservoir becomes net external output, while the remainder increases the irreducible mass. The corresponding thermodynamic efficiency and energy utilization efficiency (EUE)~\cite{hu_third_law_2026} are
\begin{equation}
\eta_{\text{thermo}}=-\frac{\Delta E_{\text{rot}}}{E_0},\qquad
\eta_{\text{EUE}}
=\frac{|E_1|/b}{|E_1|/b+\Delta M_{\text{irr}}}.
\label{eq:efficiencies}
\end{equation}
The denominator of $\eta_{\text{thermo}}$ remains the incident Killing energy, so it measures reservoir consumption per unit input. The quantity $\eta_{\text{EUE}}$ measures the fraction of the consumed reservoir converted into output. Nondecreasing area implies $0<\eta_{\text{EUE}}\le1$, with $1$ attained in the reversible limit of unchanged area. Their relation to the single-event EROI is $\eta_{\text{EROI}}/\eta_{\text{thermo}}=b\eta_{\text{EUE}}$, so the ratio equals $\eta_{\text{EUE}}$ directly only in the Kerr limit $b=1$.

For $N$ completed extraction events, energies are summed before ratios are taken. The superscript $\text{run}$ denotes the full sequence:
\begin{equation}
\begin{aligned}
\eta_{\text{EROI}}^{\text{run}}
&=\frac{\sum_n(-E_{1,n}/b_n)}{\sum_n E_{0,n}/b_n},\\
\eta_{\text{thermo}}^{\text{run}}
&=\frac{E_{\text{rot},0}-E_{\text{rot},N}}{\sum_n E_{0,n}},\\
\eta_{\text{EUE}}^{\text{run}}
&=\frac{\sum_n(-E_{1,n}/b_n)}{E_{\text{rot},0}-E_{\text{rot},N}}.
\end{aligned}
\label{eq:run-efficiencies}
\end{equation}
The sums include all events satisfying the six conditions. At each step, $b_n$ is evaluated before decay and updated with the evolving background. The cumulative net return $\eta_{\text{EROI}}^{\text{run}}$ sums mass-generator energies $E_i/b_n$ before taking the ratio, whereas the single-event $\eta_{\text{EROI}}$ is the ratio for one event. In contrast, $\eta_{\text{thermo}}^{\text{run}}$ retains the incident Killing-energy denominator of the single-event thermodynamic efficiency. Its numerator sums the decrease in the rotational reservoir over all steps and equals the initial-to-final difference, which cannot be obtained by an arithmetic mean of single-event efficiencies. Each sequence here holds $\varepsilon_0$ and $\mu_{(0)}$ fixed, so $\sum_n E_{0,n}=N\mu_{(0)}\varepsilon_0$.

By construction of Eq.~\eqref{eq:ledger}, the cumulative net output $\sum_n(-E_{1,n}/b_n)$ equals $M_0-M_N$, and reservoir consumption equals this output plus $M_{\text{irr},N}-M_{\text{irr},0}$. Thus $\eta_{\text{EROI}}^{\text{run}}$, $\eta_{\text{thermo}}^{\text{run}}$, and $\eta_{\text{EUE}}^{\text{run}}$ measure cumulative return on input, reservoir consumption per unit incident Killing energy, and utilization of the consumed reservoir, respectively. The total energy recovered also depends on how far the extraction sequence can proceed.
\section{Single and repetitive extraction results}
\label{sec:results}

%\noindent\textbf{Physical overview.}

The magnetic field can affect the gain from each decay event and the outcome of the full sequence in different ways. In the near-extremal case, the single-event net gain and cumulative return decrease as the sampled field increases, while the number of completed extraction events first increases and then decreases, producing a peak in cumulative net output. The $\delta_0=0.4$ case instead shows decreasing return, shorter sequences, and reduced output together. We compare the efficiencies, energy partition, and orbital obstructions of the two cases below to explain this difference.

\subsection{Single Penrose process efficiencies}
\label{sec:results-controls}

Both magnetic-field scans fix the initial thermodynamic mass $M_0=1$, parent specific energy $\varepsilon_0=2$, fragment angular-momentum input $\bar\ell_1=-10$, and mass ratio $\lambda=0.6$. Both the incident and output boundaries are coordinate infinity. The initial geometry is reconstructed separately at each field on the smaller-root branch continuously connected to Kerr. The spins and decay positions are listed in Table~\ref{tab:controls}. These two cases serve as representative benchmarks for the $A_4=0$-bounded near-extremal regime and the $\varepsilon_1=0$-bounded lower-spin regime respectively.

\begin{table}[tbp]
\caption{Inputs for the two magnetic-field scans. The listed condition determines $\beta_{0,\text{max}}$, the reference boundary for the initial event.}
\label{tab:controls}
\begin{ruledtabular}
\begin{tabular}{ccccc}
Case & $\delta_0$ & $r_{\text{d}}/M_0$ & $\beta_{0,\text{max}}$ & Boundary \\
\hline
Near-extremal & 0.01 & 1.2 & $\simeq0.2657$ & $A_4=0$ \\
Lower-spin & 0.4 & 1.95 & $\simeq0.07998$ & $\varepsilon_1=0$ \\
\end{tabular}
\end{ruledtabular}
\end{table}

\begin{figure*}[!t]
\centering
\includegraphics[width=\textwidth]{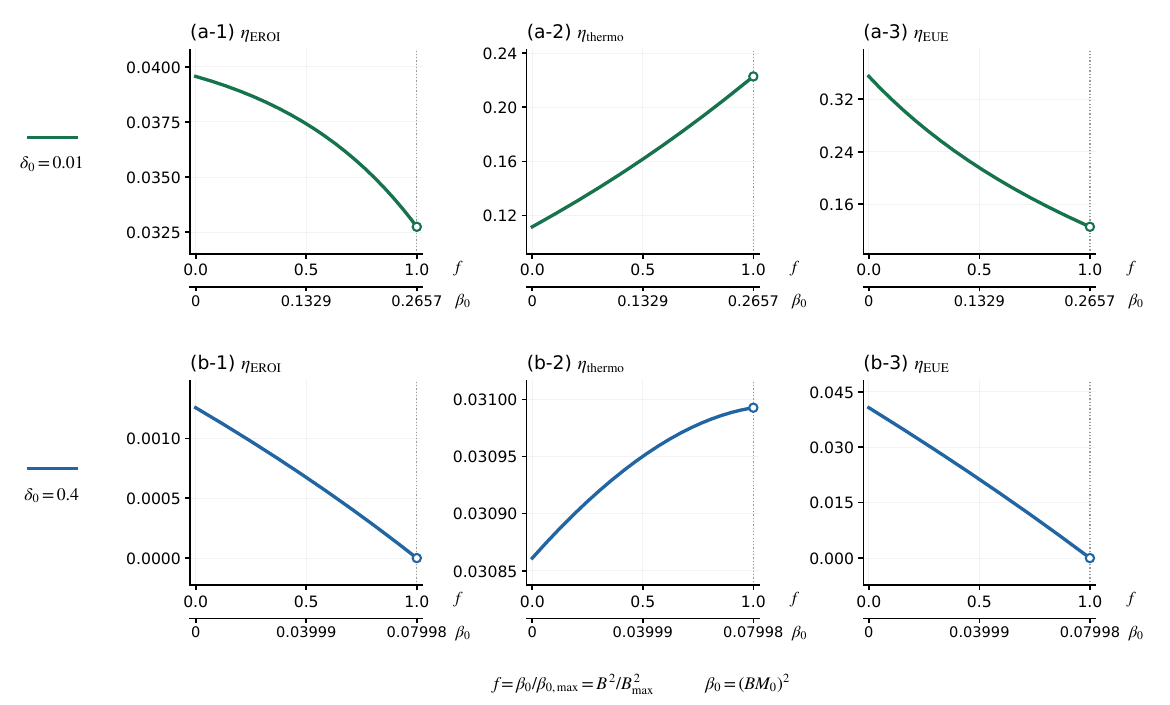}
\caption{Single-event efficiencies for (a-1)--(a-3), $\delta_0=0.01$, and (b-1)--(b-3), $\delta_0=0.4$, where $\delta_0=1-(J_0/M_0^2)^2$ and the remaining inputs are in Table~\ref{tab:controls}. Columns show $\eta_{\text{EROI}}$, $\eta_{\text{thermo}}$, and $\eta_{\text{EUE}}$ on separate linear scales to resolve their different magnitudes and trends. Each horizontal axis has two closely spaced tick rows, with $f$ above and the actual $\beta_0$ below, labeled at the right. Their full definitions appear once in the shared key below the panels. Open circles denote separately evaluated boundary limits, and the vertical dotted line marks $f=1$. The lower-spin zero-energy endpoint is excluded.}
\label{fig:spp}
\figurepanellabel{fig:spp:a-1}{a-1}
\figurepanellabel{fig:spp:a-2}{a-2}
\figurepanellabel{fig:spp:a-3}{a-3}
\figurepanellabel{fig:spp:b-1}{b-1}
\figurepanellabel{fig:spp:b-2}{b-2}
\figurepanellabel{fig:spp:b-3}{b-3}
\end{figure*}

The SPP fixes $\bar\ell_1=L_1/(M_0\mu_{(1)})$ and uses a parent mass ratio $\mu_{(0)}/M_0=10^{-6}$. In the figures, $f=\beta_0/\beta_{0,\text{max}}=B^2/B_{\text{max}}^2$ expresses the initial field as a fraction of the upper bound for each case, and the lower tick row gives the actual $\beta_0=(BM_0)^2$. Figure~\ref{fig:spp} shows the same sampled trend in both cases: $\eta_{\text{EROI}}$ and $\eta_{\text{EUE}}$ decrease, while $\eta_{\text{thermo}}$ increases. For the near-extremal case, the single-event EROI decreases from $3.956\%$ at zero field to $3.275\%$ in the reference-boundary limit, and $\eta_{\text{EUE}}$ decreases from $35.46\%$ to $12.58\%$. Meanwhile, $\eta_{\text{thermo}}$ increases from $0.1116$ to $0.2229$. More of the rotational reservoir is therefore consumed per unit incident energy, while a smaller fraction of that consumption becomes external output.

For the lower-spin case, the initial single-event EROI is approximately $0.1256\%$ and $\eta_{\text{EUE}}$ is approximately $4.069\%$. Both approach zero at the upper bound, while $\eta_{\text{thermo}}$ remains approximately $0.031$. This case reaches the fragment's zero-energy boundary first, while the parent remains accessible. Strictly negative-energy extraction ends on the inner side of that boundary. Zero net energy output can coexist with positive rotational-reservoir consumption because the counterrotating angular-momentum flux still increases the irreducible mass. The near-extremal case instead reaches $A_4=0$ first and retains nonzero net output when evaluated using the actual reduced-degree radial polynomial.

Every interior event on the single-event curves passes all six checks, and boundary limits are evaluated separately. Reducing the particle mass from $10^{-6}M_0$ to $10^{-7}M_0$ changes the thermodynamic measures by maximum relative amounts of approximately $5.81\times10^{-6}$ and $2.97\times10^{-7}$ for the two cases, respectively.

\subsection{Repetitive extraction and energy partition}
\label{sec:results-rpp}

All repetitive sequences use $\mu_{(0)}/M_0=10^{-5}$. Each sequence holds $B$, $r_{\text{d}}/M_0$, $\varepsilon_0$, and $\lambda$ fixed, and controls the fragment angular momentum using the current thermodynamic mass:
\begin{equation}
\bar\ell_1=\frac{L_{1,n}}{M_n\mu_{(1),n}}=-10,\qquad
\ell_{1,n}=\frac{M_n}{m_n}\bar\ell_1.
\label{eq:rpp-policy}
\end{equation}
Both $M_n,m_n$ are evaluated before decay at the current step. The radius $r_{\text{d}}$ and parent rest mass remain fixed relative to the original $M_0$. Eleven sequences with different initial fields are computed for each case and advanced until the next attempted event first triggers a physical obstruction, without an imposed event-count cap. The terminal state used in the statistics is always the background after the preceding completed update.

\begin{figure*}[!t]
\centering
\includegraphics[width=\textwidth]{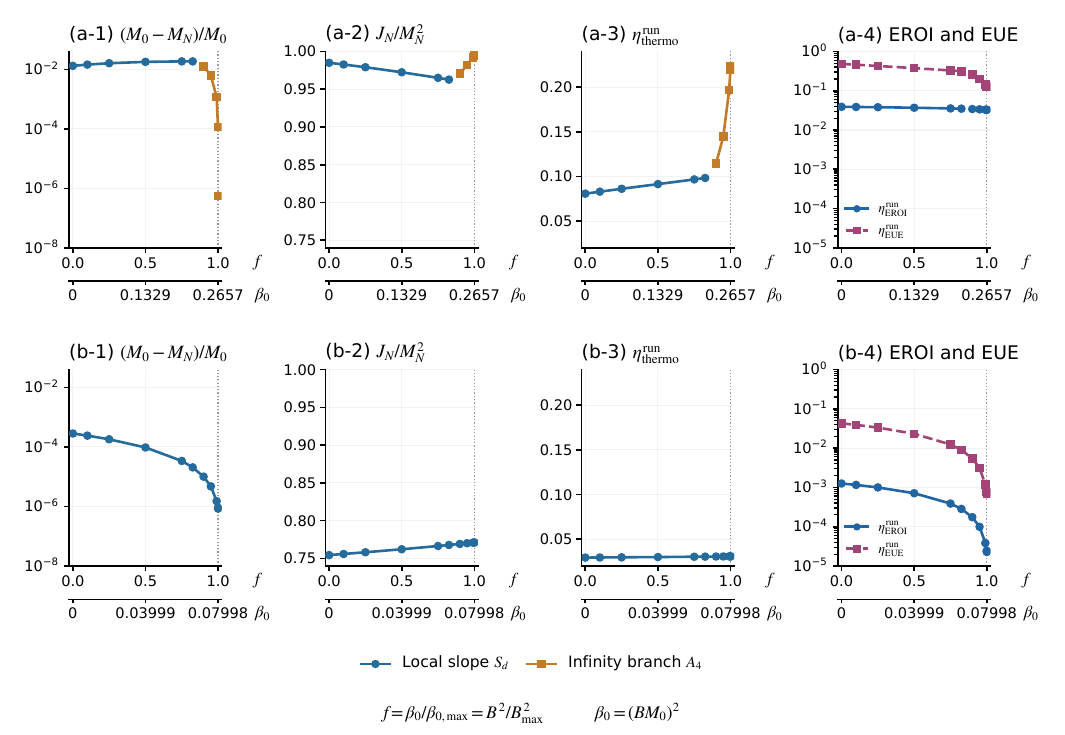}
\caption{Net energy output, final black-hole spin, and cumulative efficiencies after repetitive Penrose extraction in the near-extremal configuration, (a-1)--(a-4), and the lower-spin configuration, (b-1)--(b-4). Columns show net output, final spin, $\eta_{\text{thermo}}^{\text{run}}$, and the combined $\eta_{\text{EROI}}^{\text{run}}$ and $\eta_{\text{EUE}}^{\text{run}}$, respectively. The vertical axes in (a-1), (b-1), (a-4), and (b-4) are logarithmic, while the others are linear. Corresponding columns use the same vertical ranges. In (b-3), $\eta_{\text{thermo}}^{\text{run}}$ increases monotonically over the sampled fields, from approximately $0.02937$ at $f=0$ to $0.03070$ at $f=1-10^{-12}$ ($\beta_0\simeq0.07998$). Every horizontal axis gives $f$ in the upper tick row and $\beta_0$ in the lower row, with short labels at the right and full definitions in the shared key below the panels. In the first three columns, blue circles and orange squares denote local-slope failure and closure of the incident infinity branch. In the fourth column, blue solid circles and magenta dashed squares distinguish EROI and EUE. Curves retain breaks between termination mechanisms and isolate the near-boundary single-event sample.}
\label{fig:rpp-terminal}
\figurepanellabel{fig:rpp-terminal:a-1}{a-1}
\figurepanellabel{fig:rpp-terminal:a-2}{a-2}
\figurepanellabel{fig:rpp-terminal:a-3}{a-3}
\figurepanellabel{fig:rpp-terminal:a-4}{a-4}
\figurepanellabel{fig:rpp-terminal:b-1}{b-1}
\figurepanellabel{fig:rpp-terminal:b-2}{b-2}
\figurepanellabel{fig:rpp-terminal:b-3}{b-3}
\figurepanellabel{fig:rpp-terminal:b-4}{b-4}
\end{figure*}

Figure~\ref{fig:rpp-terminal} compares cumulative output, spin, and the three efficiencies. In the near-extremal case, from zero field to $f=0.75$, the event count increases from $16\,834$ to $29\,259$ and net output rises from approximately $1.311\%M_0$ to $1.854\%M_0$, while $\eta_{\text{EROI}}^{\text{run}}$ decreases from $3.895\%$ to $3.542\%$. The corresponding cumulative incident mass-generator energy increases from $0.3367M_0$ to $0.5236M_0$. More completed decays and a larger cumulative input offset the decline in average net return.

From the orbital perspective, each capture changes the mass and angular momentum, so the next particle attempts incidence on a new geometry. The magnetic field changes both the energy partition of an individual step and the length of the evolution path within the feasible domain. The output peak reflects their combined effect: over part of the near-extremal field range, a lower single-step return accompanies a longer extraction sequence, which means more events. At still larger fields, the large-radius transport constraint shortens the range over which extraction can continue.

The lower-spin case responds differently. From zero field to near the upper bound, the event count decreases from $11\,296$ to $1\,951$, net output decreases from approximately $2.838\times10^{-4}M_0$ to $8.475\times10^{-7}M_0$, and both $\eta_{\text{EROI}}^{\text{run}}$ and $\eta_{\text{EUE}}^{\text{run}}$ decrease. The terminal spin rises from approximately $0.7544$ to $0.7710$. For these inputs, both the average return and the sustainable event count decrease, so the sampled output peak of the near-extremal case does not occur. These sampled trends provide a baseline for investigating optimal extraction strategies across the continuous parameter space.

The cumulative thermodynamic efficiency follows a different trend from net return. Using Eq.~\eqref{eq:run-efficiencies} to divide the initial-to-final reservoir decrease directly by $N\mu_{(0)}\varepsilon_0$, the near-extremal $\eta_{\text{thermo}}^{\text{run}}$ increases from approximately $0.08067$ at zero field to $0.09664$ at $f=0.75$, reaching $0.2228$ for the single-event sample near the upper bound. The lower-spin value increases gradually from approximately $0.02937$ to $0.03070$. In both sampled cases, more of the rotational reservoir is consumed per unit incident Killing energy, while the fraction converted into output, $\eta_{\text{EUE}}^{\text{run}}$, decreases. A higher thermodynamic efficiency here therefore means greater reservoir consumption, rather than a higher net return or cumulative output.

\begin{figure*}[!t]
\centering
\includegraphics[width=\textwidth]{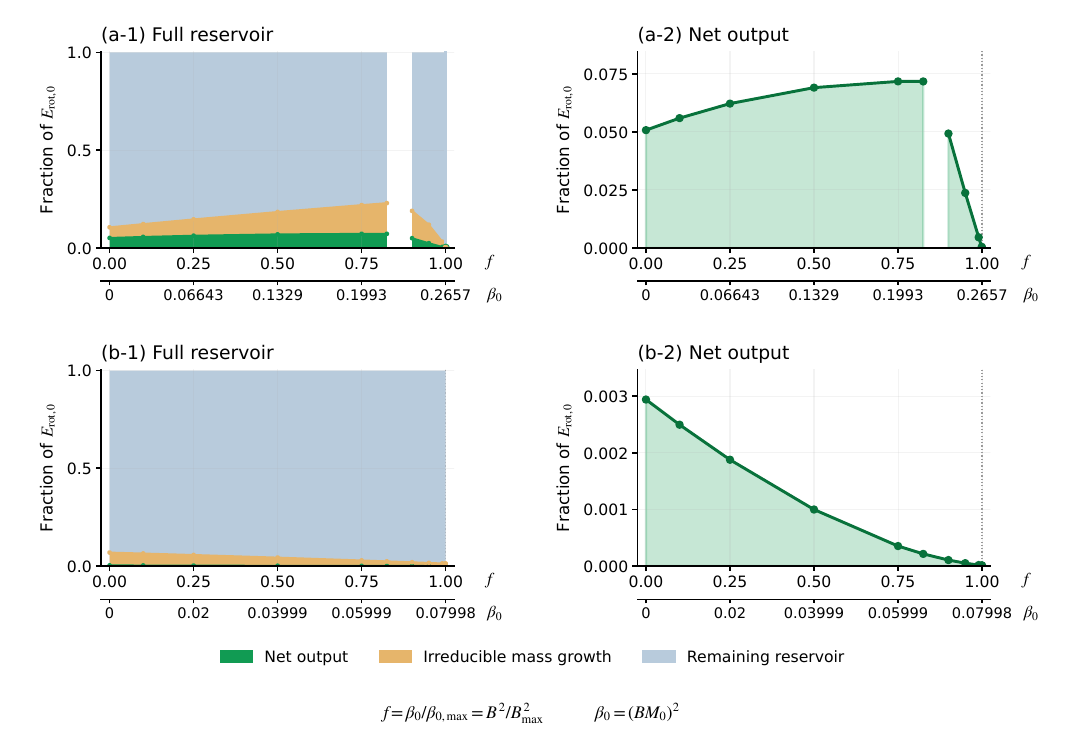}
\caption{Rotational-energy partition in the near-extremal case, (a-1) and (a-2), and the lower-spin case, (b-1) and (b-2). The first column shows the full reservoir, and the second shows net output on an enlarged linear vertical scale. All panels are normalized by their own initial $E_{\text{rot},0}$. Green, orange, and blue denote net output, irreducible-mass growth, and the remaining reservoir. The closely spaced horizontal tick rows give $f$ above and $\beta_0$ below, with short labels at the right. Their full definitions appear in the shared key below the panels. Samples with different termination mechanisms and the isolated one-event sample are left unconnected, producing the breaks in panels (a-1) and (a-2).}
\label{fig:reservoir}
\figurepanellabel{fig:reservoir:a-1}{a-1}
\figurepanellabel{fig:reservoir:a-2}{a-2}
\figurepanellabel{fig:reservoir:b-1}{b-1}
\figurepanellabel{fig:reservoir:b-2}{b-2}
\end{figure*}

Figure~\ref{fig:reservoir} places the cumulative output in the context of the initial rotational reservoir. At its sampled maximum-output point, the near-extremal case recovers only approximately $7.18\%$ of the initial reservoir; irreducible-mass growth accounts for approximately $14.71\%$, and approximately $78.11\%$ remains in the background. At zero field in the lower-spin case, net recovery is approximately $0.294\%$, irreducible-mass growth accounts for approximately $6.58\%$, and approximately $93.13\%$ remains. A separate enlarged net-output column retains the same initial-reservoir normalization so that small fractions remain visible. The denominator of $\eta_{\text{EUE}}^{\text{run}}$ is the consumed reservoir, whereas the recovered fraction of the initial reservoir uses the entire initial reservoir. These quantities answer different questions for the same sequences.

\subsection{Orbital termination mechanisms}
\label{sec:results-termination}

\begin{figure*}[!t]
\centering
\includegraphics[width=\textwidth]{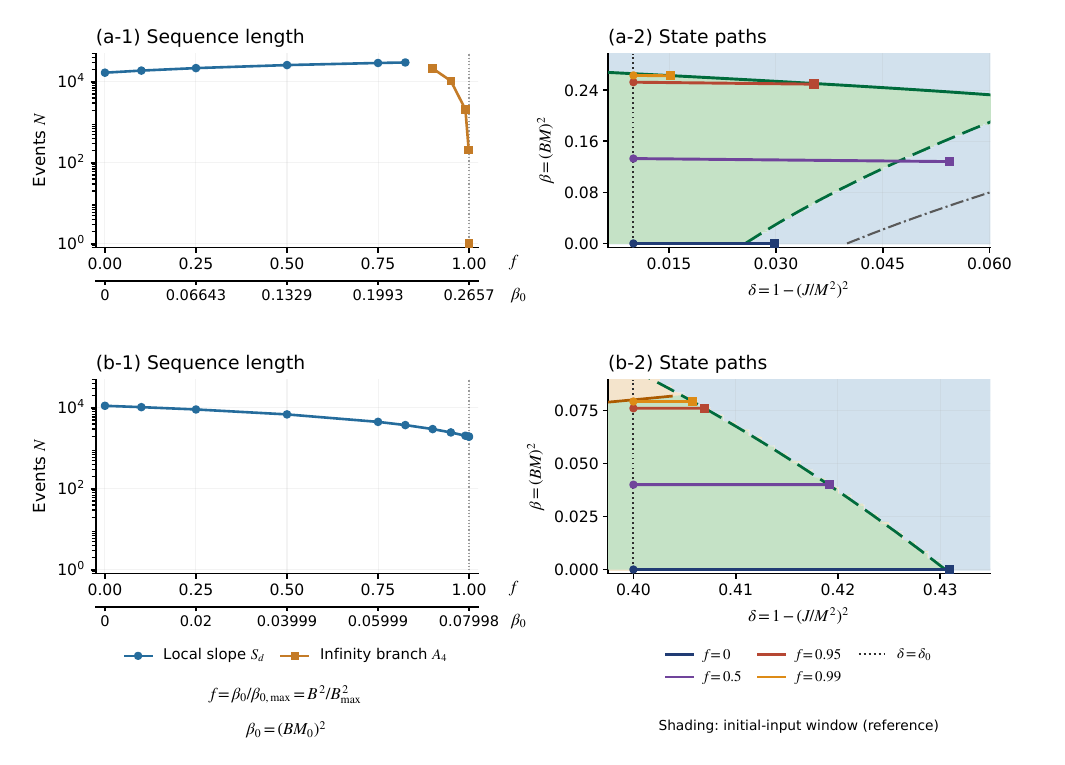}
\caption{Sequence lengths and state paths for the near-extremal case, (a-1) and (a-2), and the lower-spin case, (b-1) and (b-2). The sequence-length panels (a-1) and (b-1) use logarithmic vertical axes for the event count $N$ and two closely spaced horizontal tick rows: $f$ above and $\beta_0$ below, labeled at the right. The legend for the termination mechanisms and full field definitions are below (b-1). Both state-path panels share a single trajectory legend below (b-2). Colors and markers in the sequence-length panels identify the termination mechanisms. The state-path panels use linear axes, $\delta=1-(J/M^2)^2$ and $\beta=(BM)^2$, and display the realized paths starting at $f=0,0.5,0.95,0.99$. Circles and squares mark initial and last completed states. The background reproduces the corresponding initial-input window from Fig.~\ref{fig:window} as a reference, rather than an acceptance test along the evolving path: the fixed physical radius corresponds to a changing $r_{\text{d}}/M$. Each path panel is cropped to its local coordinate range, particularly for $\delta_0=0.4$.}
\label{fig:stop}
\figurepanellabel{fig:stop:a-1}{a-1}
\figurepanellabel{fig:stop:a-2}{a-2}
\figurepanellabel{fig:stop:b-1}{b-1}
\figurepanellabel{fig:stop:b-2}{b-2}
\end{figure*}

Figure~\ref{fig:stop} shows event counts colored by the termination mechanism and representative background trajectories for both cases. Near-extremal samples with $f\le0.825$ stop through failure of $S_{\text{d}}$, while those with $f\ge0.9$ stop through failure of $A_4$. The former means that the parent can no longer reach the prescribed turning point from its immediate exterior neighborhood, and the latter means that the allowed branch connecting that point to infinity has closed. Checks of all real roots on both sides confirm the respective obstructions, bracketing the class transition within $0.825<f<0.9$.

The output turnover and the mechanism transition can be located separately: net output already decreases from $f=0.75$ to $0.825$, while both endpoints are stopped by the same local condition. All eleven in-domain lower-spin sequences also stop through local-slope failure, with $A_4>0$ still holding for the rejected candidates. Thus, how the magnetic field changes the endpoint depends on the inputs, and the first orbital condition to fail is a different criterion from an extremum of cumulative output.

Near the initial upper bound, finite particle mass also sets the single-step resolution. At the near-extremal value $f=1-10^{-12}$, both the main-mass run and a control with ten times smaller mass complete only one extraction, with net outputs differing by a factor of ten. These are retained as finite-step samples, and the single-event point is disconnected from the curves. For the remaining ten near-extremal samples, net output differs from the $10^{-6}M_0$ runs by no more than $0.146\%$, while the sampled peak and termination mechanisms are preserved. Because these global radial obstructions halt extraction well before the rotational reservoir is depleted, varying the decay strategy offers substantial scope for further energy recovery.

% Keep the paired field-comparison figures with their discussion.
\subsection{Trajectory and endpoint dependence}
\label{sec:common-path}

%universality

To examine the evolution direction separately from input accessibility, we perform another comparison at fixed field, taking $M_0=1$, $J_0=\sqrt{0.92}$, $B=0.5$, the same initial smaller-root geometry, and $r_{\text{d}}/M_0=1.331774078416065518\ldots$. We control the fragment angular momentum at $\bar\ell_1=-19.434$ and the mass ratio at $\lambda=0.78345$, with $\mu_{(0)}/M_0=10^{-5}$. Each sequence fixes $\varepsilon_0=\kappa b_0$, where $b_0=1.166105823363214315\ldots$, for $\kappa=4.5,5,5.5$. The quantity $b_0$ always denotes its initial-state value.

\begin{figure}[!t]
\centering
\includegraphics[width=\columnwidth]{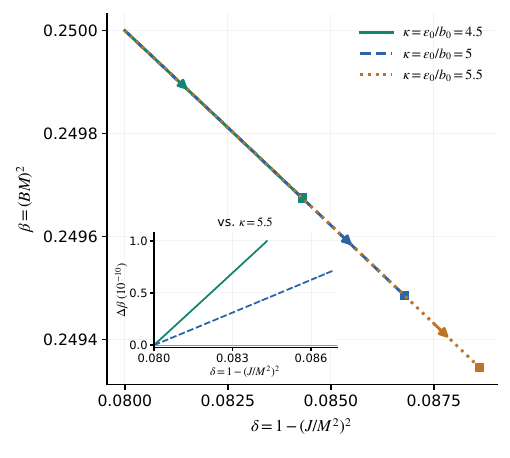}
\caption{Changing the parent specific energy moves the endpoint while preserving nearly the same state path. The legend gives $\kappa=\varepsilon_0/b_0$, and arrows indicate evolution toward the last completed states, marked by squares. The axes use $\delta=1-(J/M^2)^2$ and $\beta=(BM)^2$. The inset shows $\Delta\beta$ at the same $\delta$ relative to $\kappa=5.5$, comparing the existing piecewise-linear paths only over their common domain. This resolves the deviation between finite-mass paths.}
\label{fig:common-path}
\end{figure}

The three sequences complete $4\,066$, $7\,137$, and $9\,963$ extractions, respectively, and all stop when the parent's incident branch from infinity closes. In Fig.~\ref{fig:common-path}, the state paths nearly coincide on the main plotting scale but have different endpoints. The residual inset resolves small differences between the finite-mass point sequences. This comparison changes only the parent specific energy and is separate from the preceding field scans for two choices of spin and decay position.

This property follows from the direction of a single update. With angular-momentum input normalized by the current mass, the fragment specific energy on a given background is
\begin{equation}
\begin{gathered}
\varepsilon_1=\omega_{\text{d}} M\bar\ell_1+
N_{\text{d}}\sqrt{1+M^2\bar\ell_1^2/R_{\text{circ,d}}^2},\\
\frac{\Delta J}{\Delta M}=\frac{Mb\bar\ell_1}{\varepsilon_1}.
\end{gathered}
\label{eq:common-direction}
\end{equation}
Changing $\varepsilon_0$ changes the mass fraction $\sigma$, but this common factor cancels from the ratio of angular-momentum and mass increments. In the continuous infinitesimal-particle model, where $\mu_{(0)}\to0$ while the accumulated change remains finite,
\begin{equation}
\frac{\mathrm{d}J}{\mathrm{d}M}
=\frac{M b(M,J;B)\bar\ell_1}{
\varepsilon_1(M,J;B,r_{\text{d}},\bar\ell_1)}.
\label{eq:common-curve}
\end{equation}
When parameter mapping is smooth on the selected branch and $\varepsilon_1<0$, the right-hand side is a smooth state function independent of the parent specific energy, and local uniqueness yields a common integral curve. Incident accessibility still depends on $\varepsilon_0$, so the valid extent of each sequence along the common curve differs. Finite-rest-mass calculations instead produce discrete paths with different step sizes. The figure uses 100-digit decimal arithmetic and piecewise-linear comparisons on the union of the two node sets over their common $\delta$ interval, giving a maximum $\beta$ deviation of approximately $1.0\times10^{-10}$.

\section{Conclusions and discussion}
\label{sec:conclusion}

We have established a framework for analyzing single and repetitive Penrose processes in the genuine uncharged KBR0 background that connects particle decay, radial transport, and thermodynamic evolution. The normalization difference between coordinate Killing energy and mass-generator energy requires updating the black-hole mass by $E_1/b$. The updated thermodynamic state is then reconstructed as a background geometry by solving the cubic equation for parameter mapping on the selected branch. For neutral equatorial particles with a common radial turning point and identical incident and output boundaries, parent accessibility and future-directed negative-energy decay jointly guarantee transport of both fragments on the background. Within a valid geometry covering the exterior radial half-line and on the selected future branch, incidence from infinity further reduces to $S_{\text{d}}>0$ and $A_4\ge0$. Together with the validity of the finite final black-hole state and nondecreasing outer-horizon area, these conditions fully characterize realizable extraction events.

The numerical results show that the external magnetic field affects single-step net return, rotational-reservoir consumption, and cumulative sequence output differently. Across the field samples for both prescribed input configurations, the single-event $\eta_{\text{EROI}}$ and cumulative $\eta_{\text{EROI}}^{\text{run}}$ and $\eta_{\text{EUE}}^{\text{run}}$ decrease together as the field increases, whereas the single-event and cumulative thermodynamic efficiencies $\eta_{\text{thermo}}$ and $\eta_{\text{thermo}}^{\text{run}}$ increase. For cumulative extraction, the near-extremal case ($\delta_{0} = 0.01$) exhibits longer sequences and larger cumulative inputs over part of the field range, which offset the lower average net return and produce a sampled peak in cumulative net output. In the lower-spin case ($\delta_{0} = 0.4$), by contrast, both sequence length and cumulative output decrease with the field. These contrasting behaviors demonstrate that the magnetic-field response of cumulative extraction depends sensitively on the initial spin regime and decay radius.

The simultaneous decrease in net return and rotational energy utilization efficiency, accompanied by the increase in thermodynamic efficiency, stems from local energy partition and irreversible area growth. At fixed initial $(M_{0}, J_{0})$ and particle inputs, increasing the magnetic field modifies $N_{\text{d}}$, $\omega_{\text{d}}$ and $R_{\text{circ,d}}$ at the decay point. In both single-event scans, the positive local-energy term in Eq.~\eqref{eq:local-constants} grows with field strength, shifting the captured specific energy $\varepsilon_{1}$ closer to zero and reducing the net gain per unit incident energy after mass partition. This suppression of net energy return is qualitatively consistent with the decrease in ordinary neutral-particle Penrose efficiency with increasing magnetic field found in the weak-field regime of the magnetized Kerr spacetime by Chakraborty et al.~\cite{PhysRevD.109.064062}, motivating further study into possible common mechanisms of magnetic suppression across different background geometries and extraction setups. Beyond kinematic energy gain, however, the distinct thermodynamic efficiencies introduced here are governed by how rotational braking is partitioned into mass output and horizon entropy. Specifically, a larger normalization factor $b$ further lowers the net output measured by the mass generator and reduces the black-hole mass extracted per unit counterrotating angular momentum, $|E_{1}|/(b|L_{1}|)$. With $S$, $T$ and $\Omega$ fixed by the adopted mass relation at $(M_{0}, J_{0})$, the first law for nonextremal small steps gives $T\Delta S \simeq E_{1}/b - \Omega L_{1}$: as mass extraction drops relative to the counterrotating angular-momentum flux, more of the rotational-energy change goes into irreversible area growth, whose energy cost appears as an irreducible-mass increment $\Delta M_{\text{irr}}$. By the finite-difference budget in Eq.~\eqref{eq:reservoir-split}, reservoir consumption equals net output plus $\Delta M_{\text{irr}}$. Because the growth in $\Delta M_{\text{irr}}/E_{0}$ outpaces the drop in the net-output fraction across our samples, the fraction of consumed reservoir converted into output ($\eta_{\text{EUE}}$) decreases, whereas reservoir consumption per unit incident Killing energy ($\eta_{\text{thermo}}$) increases. While this first-order relation provides the qualitative picture, all numerical efficiencies are evaluated from exact finite area differences, and repetitive sequences accumulate this partition step by step.

Beyond step-level partition, the global evolution of repetitive sequences is governed by orbital accessibility and state-space geometry. All extraction sequences studied here terminate through loss of orbital accessibility rather than exhaustion of rotational energy, leaving terminal states in both spin regimes with most of their initial rotational reservoirs intact. As the magnetic field increases, the near-extremal termination mechanism shifts from failure of the local slope $S_{\text{d}}$ to closure of the large-radius incident branch $A_{4}$, with the cumulative net-output turnover occurring prior to this transition; all in-domain lower-spin samples terminate via the local-slope condition. Furthermore, the fixed-field comparison of parent specific energies demonstrates that, at fixed decay radius and current-mass-normalized angular momentum, varying the parent specific energy alters the mass partition and discrete step size without changing the single-step update direction on a given background. Consequently, the continuous infinitesimal-particle limit defines a common state-space integral curve, while finite-mass sequences trace nearly coincident but distinguishable paths toward different accessible endpoints.

For research on KNBR spacetimes, these results first raise the problem of optimal extraction~\cite{Bravetti:2015xsp} under simultaneous geometric, thermodynamic, and accessibility constraints. Since substantial rotational energy remains when a fixed strategy stops, can successive changes of decay radius, incident specific energy, angular momentum, and mass partition bypass the current orbital obstruction? Is the maximum recoverable energy ultimately controlled by the remaining rotational reservoir, irreversible area growth, or geometric and transport boundaries? The difficulty is that a candidate with a higher local gain may not be realizable from outside. Each strategy change requires renewed checks of global orbits and the finite final state. Near a degenerate turning point or a fold of the parameter mapping, one must also determine whether the strategy path retains finite arrival time and same-branch continuity.

The seed metrics and solution families obtained through demagnetization and remagnetization~\cite{ma_demagnetizing_2026} offer a further way to compare how different background vacua, geometric deformation, and electromagnetic coupling separately affect energy recovery. Comparing neutral and charged particles, and genuine uncharged and general charged KNBR branches, under controlled initial conditions may distinguish the effects of background geometry from those of the electromagnetic force on particles. The difficulty goes beyond adding a charge flux: different backgrounds need not share the same outer-boundary structure, and they lack a common asymptotically flat time normalization. A physically consistent interpretation of efficiency differences therefore requires first specifying which physical quantities are held fixed, which external observers receive the output, and how the energy generators of the solution families correspond.

If the external magnetic field is further treated as a physically adjustable dynamic environment, one may ask whether spin braking and external-field variation can be combined into new extraction paths, or even into repeatable energy-conversion cycles. The algebraic demagnetizing transformation used in solution generation does not itself specify a realizable time evolution. Such questions require a common energy budget that includes work by the field sources, radiative fluxes, and background changes, so that output can be attributed to black-hole rotation or to the external environment. Self-force and radiation reaction accumulated over long sequences, as well as the continuous infinitesimal-particle limit near critical obstructions, may also shift the sequence endpoints obtained from successive test-particle updates alone.

Extending particle decay to scattering and superradiance of continuous wave fields raises a broader question: does the same negative-energy-flux mechanism lead to similar long-term recovery limits in different background vacua? The challenge then shifts to wave boundary conditions and long-time evolution. Absorption, reflection, and energy return must all be specified, and a local amplification factor alone cannot determine the cumulative energy available. Along these directions, the connection between orbital accessibility, irreversible consumption, and sustained extraction can be extended from a prescribed particle extraction sequence to a common research problem for open energy systems in different electromagnetic vacua.

\acknowledgments

This work is supported by the National Natural Science Foundation of China (No.~12375059),  and the Project of National Astronomical Observatories, Chinese Academy of Sciences (No.~E4TG6601).

%the National Key Research and Development Program of China (Nos. 2021YFC2203001, 2021YFC2201901),

\appendix
\renewcommand{\thesubsection}{\arabic{subsection}}
\makeatletter
\renewcommand{\p@subsection}{\thesection.}
\makeatother
\section{Metric functions and parameter mapping}
\label{app:geometry}

In this section, we specify the metric functions, axial normalization, and equatorial frame, and state the geometric condition for a valid exterior. We then derive parameter mapping from thermodynamic states to metric parameters, classify its positive-root branches and fold, and explain how the outer-horizon area selects the physical entropy branch.

\subsection{Equatorial geometry and ZAMO frame}

The functions in Eq.~\eqref{eq:metric} are~\cite{ovcharenko_kerr-newman_2026}
\begin{equation}
\begin{gathered}
\rho^2=r^2+a^2\cos^2\theta,\quad
P=1+B^2\mu_{\text{h}}^2\cos^2\theta,\\
I=(1-mB^2r)^2+B^2r^2,\quad Q=I\Delta,\\
\Delta=(r^2+a^2)b^2-2mr,\quad
\mathcal W_{\text{c}}^2=I-B^2\Delta\cos^2\theta.
\end{gathered}
\label{eq:metric-functions}
\end{equation}
At both poles, $P=K$, so $\varphi$ has period $2\pi/K$, and the regular angle is $\phi=K\varphi$. All particle angular momenta below are $p_\phi$. The equatorial plane is a totally geodesic submanifold. Defining $D=(r^2+a^2)^2-a^2Q$, we have
\begin{equation}
\begin{gathered}
N^2=\frac{r^2\Delta}{D},\quad
\omega=\frac{aK(r^2+a^2-Q)}{D},\quad
R_{\text{circ}}^2=\frac{D}{Ir^2K^2},\\
g_{rr}=\frac{r^2}{I^2\Delta},\qquad
n=\frac{\partial_t+\omega\partial_\phi}{N},\quad
e_{\hat\phi}=\frac{\partial_\phi}{R_{\text{circ}}}.
\end{gathered}
\label{eq:zamo}
\end{equation}
Since $g_{tt}=-N^2+\omega^2R_{\text{circ}}^2$, the ergoregion condition is $N<|\omega R_{\text{circ}}|$; inside it, the frame-dragging product has the same sign as $a$. In the exterior, $Q'=2(b^2r-m)(I+B^2\Delta)>0$, with $Q(r_+)=0$ and $Q\to\infty$. Hence $Q=a^2$ has a unique exterior root when $a\ne0$.

Transport to infinity additionally requires $D>0$ throughout the exterior radial half-line. Setting $y=r-r_+$ gives the expansion
\begin{equation}
\begin{aligned}
D={}&\Gamma_\infty y^4+\frac{4(m+\mu_{\text{h}}\Gamma_\infty)}{b^2}y^3\\
&+\frac{7m^2+12m\mu_{\text{h}}+5\mu_{\text{h}}^2\Gamma_\infty}{b^4}y^2\\
&+\frac{2(4m^3+7m^2\mu_{\text{h}}+4m\mu_{\text{h}}^2+\mu_{\text{h}}^3\Gamma_\infty)}{b^6}y\\
&+\frac{4m^2(m+\mu_{\text{h}})^2}{b^8}.
\end{aligned}
\label{eq:geometry-shift}
\end{equation}
For $\Gamma_\infty\ge0$, every lower-order coefficient is strictly positive, while for $\Gamma_\infty<0$, the expression is negative at sufficiently large radius. This gives the necessary and sufficient geometric condition used in the main text, including the boundary with vanishing leading coefficient. For a finite outer boundary, only the actual interval needs to be checked.

\subsection{Parameter mapping and horizon entropy}
\label{app:parameter-mapping}

From $M=m/(Kb)$ and $J=ma/K^2$, we obtain $j=ab^2/m$ and $\delta=\mu_{\text{h}}^2/m^2$, hence $a=jm/b^2$ and $K=1+\delta B^2m^2$. The unsquared mass equation is $m=M(1+\delta B^2m^2)\sqrt{1+B^2m^2}$. Both sides are positive, so squaring is reversible and yields Eq.~\eqref{eq:parameter-mapping-cubic} without a spurious negative-mass branch.

With $u=\beta x$, the positive-root surface can be parameterized as
\begin{equation}
\begin{gathered}
x=(1+u)(1+\delta u)^2,\qquad
\beta=\frac{u}{(1+u)(1+\delta u)^2},\\
\frac{\mathrm{d}\beta}{\mathrm{d}u}
=\frac{1-\delta u-2\delta u^2}{(1+u)^2(1+\delta u)^3}.
\end{gathered}
\label{eq:parameter-mapping-parametric}
\end{equation}
For $\delta>0$, the numerator has exactly one zero at $u>0$. The function $\beta$ increases from zero to a unique maximum and then tends to zero, yielding the root classification in the main text. On this surface,
\begin{equation}
\begin{gathered}
\mathcal F_x=\frac{1-\delta u-2\delta u^2}{(1+u)(1+\delta u)},\\
\mathcal F_{xx}=-2\delta\beta^2(2+\delta+3\delta\beta x)<0.
\end{gathered}
\label{eq:root-derivatives}
\end{equation}
Thus the fold is exactly a double root. The smaller and larger roots have $\mathcal F_x>0$ and $\mathcal F_x<0$, respectively. The implicit-function theorem guarantees local continuation on the same branch away from the fold, but does not authorize arbitrary finite jumps between branches.

Degenerate boundaries must be handled directly: $\beta=0$ has the unique root $x=1$; at $\delta=0$, $x=(1-\beta)^{-1}$ is finite and positive only for $0\le\beta<1$; the nonrotating boundary $\delta=1$ remains cubic, with fold $(\beta_*,x_*)=(4/27,27/8)$. Positive-root classification alone does not establish exterior validity or acceptance of a complete process.

The area formula directly gives
\begin{equation}
\begin{gathered}
S=\frac{2\pi m(m+\mu_{\text{h}})}{K^2b^2}
=2\pi M^2(1+\mu_{\text{h}}/m),\\
M^4-J^2=M^4\mu_{\text{h}}^2/m^2.
\end{gathered}
\label{eq:area-root}
\end{equation}
The condition $\mu_{\text{h}}\ge0$ selects the larger of the two entropy values obtained from the Christodoulou--Ruffini relation, corresponding to the outer horizon. Different geometric roots share this value. The two values coincide at extremality. At $J=0$, the solution with zero entropy lies outside the domain $S>0$ of the original mass formula.
\section{Particle decay and parent accessibility}
\label{app:local}

In this section, we obtain the particle rapidities, the future-directed mass partition, and the conserved quantities for particle decay. We then analyze the radial polynomial to formulate the parent-accessibility conditions and explain how root multiplicity and continuity determine finite-time arrival and preservation of the future branch.

\subsection{Particle decay kinematics}

At the decay point inside the ergoregion, $|\omega_{\text{d}} R_{\text{circ,d}}|>N_{\text{d}}$, and the parent's real future-directed rapidity is unique:
\begin{equation}
\begin{gathered}
\chi_0=\operatorname{sgn}(a)\ln\!\left[
\frac{\varepsilon_0+\sqrt{\varepsilon_0^2+\omega_{\text{d}}^2R_{\text{circ,d}}^2-N_{\text{d}}^2}}
{N_{\text{d}}+|\omega_{\text{d}} R_{\text{circ,d}}|}\right],\\
\chi_1=\operatorname{arsinh}(m\ell_1/R_{\text{circ,d}}).
\end{gathered}
\label{eq:local-rapidities}
\end{equation}
Four-momentum conservation and $g(u_2,u_2)=-1$ give the mass-partition equation
\begin{equation}
(1-\lambda^2)\sigma^2-2\Gamma_{01}\sigma+1=0.
\label{eq:mass-quadratic}
\end{equation}
In the parent rest frame, $E_{2|0}=\mu_{(0)}(1-\sigma\Gamma_{01})>0$. Only the root in Eq.~\eqref{eq:mass-ratio} is future directed, and it satisfies $\sigma(1+\lambda)\le1$. The conserved quantities are
\begin{equation}
\begin{aligned}
\ell_0={}&\frac{R_{\text{circ,d}}}{m}
\\ &\times\frac{\omega_{\text{d}} R_{\text{circ,d}}\varepsilon_0-\operatorname{sgn}(a)N_{\text{d}}
\sqrt{\varepsilon_0^2+\omega_{\text{d}}^2R_{\text{circ,d}}^2-N_{\text{d}}^2}}
{\omega_{\text{d}}^2R_{\text{circ,d}}^2-N_{\text{d}}^2},\\
\varepsilon_1={}&\omega_{\text{d}} m\ell_1+N_{\text{d}}\sqrt{1+m^2\ell_1^2/R_{\text{circ,d}}^2},\\
\varepsilon_2={}&\frac{\varepsilon_0-\sigma\varepsilon_1}{\lambda\sigma},\qquad
\ell_2=\frac{\ell_0-\sigma\ell_1}{\lambda\sigma}.
\end{aligned}
\label{eq:local-constants}
\end{equation}
Negative energy requires $N_{\text{d}}+\omega_{\text{d}} R_{\text{circ,d}}\tanh\chi_1<0$. A massive particle cannot satisfy the negative-energy condition on the stationary-limit surface itself.

\subsection{Parent radial accessibility}
\label{app:radial}

Using the dimensionless radius $\varrho=r/M_0$ and radial function $\overline{\mathcal R}=\hat{\mathcal R}/M_0^4$ of the main text, write $\varrho_{\text{d}}=r_{\text{d}}/M_0$ and $\varrho_+=r_+/M_0$. The common-turning-point condition gives
\begin{equation}
\overline{\mathcal R}(\varrho)
=(\varrho-\varrho_{\text{d}})\mathcal C_{\text{d}}(\varrho).
\label{eq:radial-factor}
\end{equation}
Here $\mathcal C_{\text{d}}$ is a radial quotient of degree at most three, whose sign determines the allowed radial region outside the turning point. The quartic and cubic coefficients of $\overline{\mathcal R}$ are expressed directly in terms of geometric quantities and particle constants:
\begin{equation}
\begin{aligned}
A_4&=\varepsilon^2-b^2-B^2b^4(a\varepsilon-Km\ell)^2,\\
A_3&=\frac{2m}{M_0}[1+2B^2b^2(a\varepsilon-Km\ell)^2]>0.
\end{aligned}
\label{eq:radial-leading}
\end{equation}
Outside the decay point, use $z=\varrho-\varrho_{\text{d}}\ge0$ as the radial displacement. Since $\overline{\mathcal R}(\varrho_+)\ge0$, the derivative of the radial quotient can be written as
\begin{equation}
\begin{aligned}
\mathcal C_{\text{d}}'(\varrho_{\text{d}}+z)
={}&\frac{S_{\text{d}}}{\varrho_{\text{d}}-\varrho_+}
+\frac{\overline{\mathcal R}(\varrho_+)}{(\varrho_{\text{d}}-\varrho_+)^2}\\
&+(\varrho_{\text{d}}-\varrho_+)[A_3+(3\varrho_{\text{d}}+\varrho_+)A_4]\\
&+2(A_3+4\varrho_{\text{d}}A_4)z+3A_4z^2.
\end{aligned}
\label{eq:radial-shift}
\end{equation}
If $S_{\text{d}}>0$ and $A_4\ge0$, the right-hand side is strictly positive for all $z\ge0$, and $\mathcal C_{\text{d}}(\varrho_{\text{d}})=S_{\text{d}}>0$, hence no exterior radial barrier exists. When $A_4=0$, the radial polynomial reduces to a cubic with positive leading coefficient, and the conclusion remains valid. Conversely, $S_{\text{d}}<0$ forbids motion immediately outside the turning point, while $A_4<0$ forbids motion at sufficiently large radius. This establishes Eq.~\eqref{eq:parent-access}. Near sign boundaries, precision must be increased or the actual reduced degree used. A nonzero negative leading coefficient cannot be discarded by a tolerance.

The future branch must also persist along the connected interval. The mass shell gives
\begin{equation}
\frac{(\varepsilon-\omega m\ell)^2}{N^2}
=1+\frac{m^2\ell^2}{R_{\text{circ}}^2}+g_{rr}(u^r)^2>0.
\label{eq:future-continuity}
\end{equation}
Within a geometrically valid, radially allowed interval, $\varepsilon-\omega m\ell$ cannot pass through zero, so the future branch selected at $r_{\text{d}}$ persists by continuity. A simple root also guarantees arrival in finite time: near a finite regular root $s$ of multiplicity $\nu$, $|\mathrm{d}\tau|\sim\text{const}\,|r-s|^{-\nu/2}|\mathrm{d}r|$, and the integral converges only for $\nu=1$. A double root therefore separates transport intervals even if the radial function does not change sign. The strict condition $S_{\text{d}}>0$ is consequently not a numerical convention that can be relaxed arbitrarily.

Infinity here denotes a connected allowed branch extending to arbitrarily large finite radii, and no asymptotically flat interpretation of energy is imposed. For a finite outer boundary, only the corresponding finite interval is checked.

\section{Proofs of global fragment transport}
\label{app:transport}

In this section, we establish the radial transport of both fragments from the common turning point. We show that a future-directed negative-energy fragment reaches the future horizon, and use a comparison photon and concavity in the impact parameter to show that parent accessibility guarantees outward transport of the other fragment to the same boundary.

\subsection{Negative-energy fragment capture}

Let $\varepsilon_1<0$ and select the future turning branch at $r_{\text{d}}$ inside the ergoregion. Denote the combination appearing in the radial function by $A_1=(r^2+a^2)\varepsilon_1-aKm\ell_1$. The subscript $\text{d}$ indicates evaluation at $r_{\text{d}}$. Future orientation and $0<Q_{\text{d}}<a^2$ give
\begin{equation}
\begin{aligned}
A_1(r_{\text{d}})&=\frac{D_{\text{d}}(\varepsilon_1-\omega_{\text{d}}m\ell_1)
-r_{\text{d}}^2Q_{\text{d}}\varepsilon_1}{r_{\text{d}}^2+a^2-Q_{\text{d}}}>0,\\
A_1(r)&=A_1(r_{\text{d}})+(-\varepsilon_1)(r_{\text{d}}^2-r^2)>0
\quad(r_+\le r<r_{\text{d}}).
\end{aligned}
\label{eq:capture-numerator}
\end{equation}
Since $Q'>0$ and $(\Delta r^2)'>0$,
\begin{equation}
\begin{aligned}
\hat{\mathcal R}_1'={}&4r\varepsilon_1A_1-Q'(a\varepsilon_1-Km\ell_1)^2\\
&-(\Delta r^2)'<0\qquad(r_+<r\le r_{\text{d}}).
\end{aligned}
\label{eq:capture-monotonicity}
\end{equation}
Starting inward from $\hat{\mathcal R}_1(r_{\text{d}})=0$, the radial function is strictly positive, and the turning root is simple. On this interval, $Q<Q_{\text{d}}<a^2$ ensures $D>0$, and Eq.~\eqref{eq:future-continuity} preserves the future branch. At the horizon,
\begin{equation}
\varepsilon_1-\Omega_{\text{h}}^{(\phi)}m\ell_1
=\frac{A_1(r_+)}{r_+^2+a^2}>0,\qquad
\Omega_{\text{h}}^{(\phi)}=\frac{aK}{r_+^2+a^2}.
\label{eq:future-capture}
\end{equation}
This also identifies the crossed horizon as the future horizon. A negative-energy turning fragment is therefore not blocked by an additional inner barrier. If its energy becomes nonnegative, the negative-energy condition has failed, which is not a capture barrier.

\subsection{Outgoing fragment escape}

First take $a>0$. Local conservation of energy and azimuthal momentum gives
\begin{equation}
v_1<v_E<v_0<v_2<1,\qquad v_E=-N_{\text{d}}/(\omega_{\text{d}}R_{\text{circ,d}}).
\label{eq:speed-order}
\end{equation}
For a positive-energy particle, introduce the shifted impact parameter $b_{\text{imp}}=Km\ell/\varepsilon-a$, which has dimensions of length. Then
\begin{equation}
\begin{gathered}
b_{\text{imp}}(v)=\frac{KR_{\text{circ,d}}v}{N_{\text{d}}+\omega_{\text{d}}R_{\text{circ,d}}v}-a,\qquad
\frac{\mathrm{d}b_{\text{imp}}}{\mathrm{d}v}>0,\\
b_{\text{imp},0}<b_{\text{imp},2}<b_{\text{imp},\gamma},
\qquad b_{\text{imp},\gamma}=\frac{r_{\text{d}}^2}{a+\sqrt{Q_{\text{d}}}}.
\end{gathered}
\label{eq:impact-order}
\end{equation}
Here $b_{\text{imp},\gamma}$ denotes a prograde comparison photon turning at the same point. Eliminating $\varepsilon^{-2}$ for a positive-energy particle using the common-turning-point condition, define
\begin{equation}
\begin{aligned}
\mathcal P_{\text{d}}(r,b_{\text{imp}})
={}&(r^2-ab_{\text{imp}})^2-Qb_{\text{imp}}^2\\
&-\frac{\Delta r^2}{\Delta_{\text{d}}r_{\text{d}}^2}
[(r_{\text{d}}^2-ab_{\text{imp}})^2-Q_{\text{d}}b_{\text{imp}}^2].
\end{aligned}
\label{eq:comparison-polynomial}
\end{equation}
For $i=0,2$, $\mathcal P_{\text{d}}(r,b_{\text{imp},i})=\hat{\mathcal R}_i/\varepsilon_i^2$. Its quadratic coefficient in $b_{\text{imp}}$ and the derivative of that coefficient at the turning point are
\begin{equation}
\begin{aligned}
A_{\text{d}}(r)&=-(Q-Q_{\text{d}})
-\frac{\Delta r^2-\Delta_{\text{d}}r_{\text{d}}^2}{\Delta_{\text{d}}r_{\text{d}}^2}(a^2-Q_{\text{d}})<0,\\
A_{\text{d}}'(r_{\text{d}})&=-Q_{\text{d}}'-\left(\frac{\Delta_{\text{d}}'}{\Delta_{\text{d}}}+\frac2{r_{\text{d}}}\right)(a^2-Q_{\text{d}})<0.
\end{aligned}
\label{eq:comparison-concavity}
\end{equation}
Thus both the radial function and its turning-point slope are strictly concave in the impact parameter.

Accessibility of the comparison photon must still be proved, and concavity alone cannot justify omitting this endpoint. Write $k_\gamma=\partial_r\mathcal P_{\text{d}}(r_{\text{d}},b_{\text{imp},\gamma})$. At the photon endpoint,
\begin{equation}
\begin{aligned}
\partial_r\partial_{b_{\text{imp}}}\mathcal P_{\text{d}}
={}&4r_{\text{d}}(a+\sqrt{Q_{\text{d}}})\frac{I_{\text{d}}-B^2\Delta_{\text{d}}}{I_{\text{d}}+B^2\Delta_{\text{d}}}\\
&-\frac{2(a\sqrt{Q_{\text{d}}}-B^2\Delta_{\text{d}}^2)}
{(I_{\text{d}}+B^2\Delta_{\text{d}})\Delta_{\text{d}}b_{\text{imp},\gamma}}k_\gamma.
\end{aligned}
\label{eq:photon-local}
\end{equation}
We have $I-B^2\Delta=1-a^2B^2b^2>0$ and $a\sqrt{Q_{\text{d}}}>Q_{\text{d}}>B^2\Delta_{\text{d}}^2$. If $k_\gamma\le0$, Eq.~\eqref{eq:photon-local} is strictly positive. An exact quadratic expansion of $\partial_r\mathcal P_{\text{d}}$ about $b_{\text{imp},\gamma}$, together with $b_{\text{imp},0}-b_{\text{imp},\gamma}<0$ and $A_{\text{d}}'(r_{\text{d}})<0$, would then make the parent's slope negative, a contradiction. Hence $k_\gamma>0$.

Suppose that the photon has a first exterior zero $r_b$ within the parent's accessible interval. Since
\begin{equation}
\mathcal P_{\text{d}}(r,b_{\text{imp},\gamma})
=[r^2-(a+\sqrt Q)b_{\text{imp},\gamma}]
[r^2-(a-\sqrt Q)b_{\text{imp},\gamma}],
\label{eq:photon-factor}
\end{equation}
the second factor is always positive, this root satisfies $r_b^2=b_{\text{imp},\gamma}(a+\sqrt{Q_b})$. At that point,
\begin{equation}
\begin{gathered}
\partial_{b_{\text{imp}}}\mathcal P_{\text{d}}(r_b,b_{\text{imp},\gamma})
=2r_b^2\left(\frac{\Delta_b\sqrt{Q_{\text{d}}}}{\Delta_{\text{d}}}-\sqrt{Q_b}\right)>0,\\
\Delta_b^2Q_{\text{d}}-\Delta_{\text{d}}^2Q_b
=\Delta_b\Delta_{\text{d}}(1-a^2B^2b^2)(\Delta_b-\Delta_{\text{d}})>0.
\end{gathered}
\label{eq:photon-barrier}
\end{equation}
Another exact quadratic expansion, using the negative difference $b_{\text{imp},0}-b_{\text{imp},\gamma}$ and $A_{\text{d}}(r_b)<0$, gives $\mathcal P_{\text{d}}(r_b,b_{\text{imp},0})<0$, contradicting parent accessibility. The photon is therefore strictly allowed over the full interval. Finally, $b_{\text{imp},2}$ lies between the parent and photon values, so the concavity in Eq.~\eqref{eq:comparison-concavity} makes both its radial function and its initial slope strictly positive. Its future branch persists by continuity.

Applying this argument to any common finite outer boundary, and then to arbitrarily large finite radii, yields the infinity result in the main text. Negative spin follows by the axial reflection of Appendix~\ref{app:spin-reflection}, with unchanged radial criteria. The proof does not select between the smaller-root and larger-root branches used for parameter mapping, but always requires the path to remain within valid geometry.
\section{Analytical decay domain constraints}
\label{app:window}

In this section, we identify the geometric decay window and the restrictions from parent accessibility and fragment negative energy. We discuss the boundary with a vanishing quartic coefficient, obtain a uniform negative-energy bound for the near-extremal input, and derive a necessary lower spin bound.

\subsection{Feasible window boundaries}

We use the geometric quantities $(m,a,B)$, thermodynamic quantities $(M_0,J_0)$, and dimensionless combinations $j_0=J_0/M_0^2$, $\delta_0=1-j_0^2$, $\beta_0=(BM_0)^2$, and $u=B^2m^2$. At fixed $(M_0,j_0)$,
\begin{equation}
\begin{gathered}
\beta_0=\frac{u}{(1+u)(1+\delta_0u)^2},\qquad
\frac{m}{M_0}=(1+\delta_0u)\sqrt{1+u},\\
\frac{a}{M_0}=\frac{j_0(1+\delta_0u)}{\sqrt{1+u}},\qquad
\Gamma_\infty=1-(1-\delta_0)u.
\end{gathered}
\label{eq:window-parameters}
\end{equation}
The horizon and stationary limit, expressed in the geometric radial coordinate, are
\begin{equation}
\begin{gathered}
 r_+=\frac{m}{b^2}(1+\sqrt{\delta_0}),\\
 r_{\text{ergo}}=\frac{m}{b^2}\left[1+
 \sqrt{\frac{2}{1-\delta_0u+\sqrt{(1-\delta_0u)^2+4u}}}\right].
\end{gathered}
\label{eq:geometric-window}
\end{equation}
The stationary limit is the exterior root of $Q(r_{\text{ergo}})=a^2$, and the expression above is regular at $u=0$. The geometric window requires only $r_+<r_{\text{d}}<r_{\text{ergo}}$, and particle conditions further restrict it.

The parent's radial quartic coefficient retains the geometric expression of Appendix~\ref{app:radial}:
\begin{equation}
A_4=\varepsilon_0^2-b^2-B^2b^4(a\varepsilon_0-Km\ell_0)^2.
\label{eq:window-A4}
\end{equation}
This gives the necessary condition $u\le\varepsilon_0^2-1$. Requiring $A_4>0$ makes it strict: $u<\varepsilon_0^2-1$. At $A_4=0$, the cubic coefficient is
\begin{equation}
A_3=\frac{2m}{M_0}[1+2B^2b^2(a\varepsilon_0-Km\ell_0)^2]>0,
\label{eq:window-A3}
\end{equation}
The equality boundary therefore cannot be rejected automatically. Nor can the necessary energy bound replace the angular-momentum term or the strictly positive slope condition.

The negative-energy fragment imposes a further radial restriction. Fix the counterrotating input $\bar\ell_1=L_1/(M_0\mu_{(1)})$ and retain $j_0\bar\ell_1<0$ inside the ergoregion. Then
\begin{equation}
\bar\ell_1^2>
\frac{r_{\text{d}}^2\Delta_{\text{d}}}{M_0^2K^2[a^2-Q_{\text{d}}]}
\label{eq:negative-threshold}
\end{equation}
is equivalent to $\varepsilon_1<0$. The outer boundary of the negative-energy window satisfies
\begin{equation}
\bar\ell_1^2M_0^2K^2[a^2-Q_{\text{d}}]-r_{\text{d}}^2\Delta_{\text{d}}=0.
\label{eq:negative-window-edge}
\end{equation}
Both $Q$ and $r^2\Delta$ increase strictly in the exterior. The left-hand side is therefore positive at the horizon, negative at the stationary limit, and strictly decreasing between them, giving a unique upper-end root. Actual candidates lie in the intersection of this negative-energy window and the parent's accessible region.

The fragment input in the upper plots of Fig.~\ref{fig:window} admits a uniform negative-energy guarantee. For a concise bound, temporarily use the normalized displacement from the horizon center, $w=b^2r_{\text{d}}/m-1$. The conditions $\Gamma_\infty\ge0$ and $r_{\text{d}}/M_0=6/5>r_+/M_0$ give $[m/(M_0b^2)]^2\ge1/2$ and $\sqrt{\delta_0}<w\le6\sqrt2/5-1$. Substituting these bounds into Eq.~\eqref{eq:negative-threshold}, its right-hand side is no greater than
\begin{equation}
\begin{gathered}
\frac{(6\sqrt2/5)^2(6\sqrt2/5-1)^2}
{1-(6\sqrt2/5-1)^2-(6\sqrt2/5-1)^4}\\
=5.0331846027\ldots<\bar\ell_1^2=100.
\end{gathered}
\label{eq:uniform-negative}
\end{equation}
Thus the negative-energy condition does not further restrict the parent-candidate domain for the near-extremal configuration in the upper plots, and $\lambda=0.6$ changes only the future-directed mass partition. This guarantee does not apply to the lower configuration with $r_{\text{d}}/M_0=1.95$.

\subsection{Necessary lower spin bound}

Take $j_0>0$ and continue to use $w=b^2r_{\text{d}}/m-1$, with $\sqrt{\delta_0}<w<1$ inside the ergoregion. The comparison photon in Appendix~\ref{app:transport} must have a positive outward slope. Using $m/b^2$ as the length unit, its sign is that of
\begin{equation}
\begin{aligned}
\mathcal K_\gamma={}&4j_0\sqrt{(1+uw^2)(w^2-\delta_0)}\\
&+4(1+uw^2)(w^2-\delta_0)\\
&-(1+w)\partial_w[(1+uw^2)(w^2-\delta_0)].
\end{aligned}
\label{eq:spin-K}
\end{equation}
At fixed $(\delta_0,w)$, its field derivative is
\begin{equation}
\begin{aligned}
\partial_u\mathcal K_\gamma
&=2w\left[\frac{j_0w(w^2-\delta_0)}
{\sqrt{(1+uw^2)(w^2-\delta_0)}}-2w^2-\delta_0(w-1)\right]\\
&\le-2w^4<0.
\end{aligned}
\label{eq:spin-field-bound}
\end{equation}
The inequality uses $(w^2-\delta_0)/\sqrt{(1+uw^2)(w^2-\delta_0)}\le w$, $j_0\le1$, and $2w^2+\delta_0(w-1)\ge w^2(1+w)$. At zero field,
\begin{equation}
\mathcal K_\gamma(0,w)
=4j_0\sqrt{w^2-\delta_0}+2w^2-2w-4\delta_0
\label{eq:spin-zero-field}
\end{equation}
is a strictly increasing function of $w$, as follows from the positive derivative $4j_0w/\sqrt{w^2-\delta_0}+4w-2>0$ together with $j_0+w>1$. Hence
\begin{equation}
0<\mathcal K_\gamma(u,w)
\le\mathcal K_\gamma(0,w)<\mathcal K_\gamma(0,1)=4-8\delta_0.
\label{eq:spin-bound}
\end{equation}
Necessarily, $\delta_0<1/2$. Axial reflection from Appendix~\ref{app:spin-reflection} extends the result to negative spin, giving $|j_0|>1/\sqrt2$. This is a necessary condition: it guarantees neither global incidence nor negative-energy decay nor final-state acceptance.
\section{Spin-reversal symmetry}
\label{app:spin-reflection}

In this section, we analyze spin reversal through axial reflection, following its action on geometry, particle decay, radial transport, thermodynamic evolution, branches of the parameter mapping, and extraction efficiencies. We also clarify the transformation of the electromagnetic field and the scope of the neutral-particle correspondence.

The particles' azimuthal motion must reverse together with the spin. At fixed $m$ and signed $B$, define
\begin{equation}
(a,\phi,J,L_i,\chi_i)\longmapsto(-a,-\phi,-J,-L_i,-\chi_i).
\label{eq:spin-reflection-map}
\end{equation}
Particle rest masses, Killing energies, and the decay radius are unchanged.

\subsection{Geometry and time orientation}

The quantities $b$, $\mu_{\text{h}}$, $K$, $I$, $\Delta$, $Q$, $P$, $\rho^2$, and $\mathcal W_{\text{c}}^2$ in Eqs.~\eqref{eq:background} and~\eqref{eq:metric-functions} are even in $a$. Thus $g_{tt}$, $g_{rr}$, $g_{\theta\theta}$, and $g_{\phi\phi}$ are even, while only $g_{t\phi}$ is odd. Let
\begin{equation}
\begin{gathered}
\mathrm{P}(t,r,\theta,\phi)=(t,r,\theta,-\phi),\\
\mathrm{P}^*g(m,-a,B)=g(m,a,B).
\end{gathered}
\label{eq:spin-metric-reflection}
\end{equation}
Here $\mathrm{P}^*$ denotes pullback. Reversal of $\mathrm{d}\phi$ cancels the sign reversal of the cross component, yielding an isometry between the two parameter-labeled geometries. The axial factor $K$ and the regular angular period $2\pi$ are unchanged, as are the horizons, stationary limit, and valid exterior domain.

Taking the positive roots for $N,R_{\text{circ}}$ in Eq.~\eqref{eq:zamo}, we obtain
\begin{equation}
\begin{gathered}
N(-a)=N(a),\quad R_{\text{circ}}(-a)=R_{\text{circ}}(a),\\
\omega(-a)=-\omega(a),\\
\mathrm{P}_*n_a=n_{-a},\qquad
\mathrm{P}_*e_{\hat\phi,a}=-e_{\hat\phi,-a}.
\end{gathered}
\label{eq:spin-frame-reflection}
\end{equation}
Subscripts label the backgrounds, and $\mathrm{P}_*$ denotes pushforward. The future ZAMO normal maps to the future normal, so this axial reflection preserves time orientation.

\subsection{Decay and negative energy}

Tildes denote reflected quantities, distinguished from the primes used for updated states. The reflected momentum one-forms satisfy $\tilde{p}_{(i)}(\mathrm{P}_*v)=p_{(i)}(v)$. Pairing them with the two Killing generators and using Eq.~\eqref{eq:spin-frame-reflection} gives
\begin{equation}
\begin{gathered}
\widetilde E_i=E_i,\quad \widetilde L_i=-L_i,\quad
\widetilde\ell_i=-\ell_i,\\
\widetilde\mu_{(i)}=\mu_{(i)},\qquad \widetilde\chi_i=-\chi_i.
\end{gathered}
\label{eq:spin-particle-reflection}
\end{equation}
The normal component $\cosh\chi_i$ of the four-velocity is unchanged, while its azimuthal component $\sinh\chi_i$ reverses sign. Reversing rapidity reverses only the local azimuthal velocity, while the real parent solution selected by Eq.~\eqref{eq:local-rapidities} also changes sign exactly. The relative Lorentz factor $\Gamma_{01}=\cosh(\chi_0-\chi_1)$ is invariant, so the future-directed mass partition $\sigma$, the ratio $\lambda$, and both fragment rest masses are unchanged. Four-momentum conservation is preserved under the isometry.

At the decay point inside the ergoregion, $\operatorname{sgn}(\omega_{\text{d}} R_{\text{circ,d}})=\operatorname{sgn}(a)$ and $|\omega_{\text{d}} R_{\text{circ,d}}|>N_{\text{d}}$. The necessary and sufficient negative-energy condition can be written uniformly as
\begin{equation}
\operatorname{sgn}(a)\chi_1<-
\operatorname{artanh}\!\left(\frac{N_{\text{d}}}{|\omega_{\text{d}} R_{\text{circ,d}}|}\right).
\label{eq:spin-negative-energy}
\end{equation}
It is invariant under reflection and requires $aL_1<0$. For negative spin, the energy-extracting fragment therefore has positive angular momentum. Both spin orientations are braked by counterrotating angular-momentum flux.

\subsection{Radial transport}

Under the simultaneous changes $a\to-a$ and $\ell\to-\ell$, the combination $(r^2+a^2)\varepsilon-aKm\ell$ in Eq.~\eqref{eq:radial} is unchanged, while $a\varepsilon-Km\ell$ changes sign and is then squared. Thus, with subscripts labeling the backgrounds,
\begin{equation}
\hat{\mathcal R}_{-a}(r;\varepsilon,-\ell)
=\hat{\mathcal R}_a(r;\varepsilon,\ell).
\label{eq:spin-radial-reflection}
\end{equation}
Radial zeros, their multiplicities, derivative signs, and allowed intervals are identical. The quantities $g_{rr}$ and $u^r$ for the same radial direction are also unchanged. The parent's $A_4,S_{\text{d}}$, full-interval connectivity of the fragments, and proper-time arrival conditions at the turning point therefore correspond one to one. Along the allowed interval, the future-branch combination $E-\omega L$ is invariant.

The horizon angular velocity associated with the regular angle is $\Omega_{\text{h}}^{(\phi)}=aK/(r_+^2+a^2)$, which is odd in $a$. Hence
\begin{equation}
E_1-\Omega_{\text{h}}^{(\phi)}(-a)(-L_1)
=E_1-\Omega_{\text{h}}^{(\phi)}(a)L_1.
\label{eq:spin-horizon-reflection}
\end{equation}
The future-horizon crossing condition is therefore preserved. The reflected image of a globally accepted orbit has the same capture or outward-transport properties, extending the established transport result to the opposite spin orientation.

\subsection{Parameter mapping and efficiencies}

Equation~\eqref{eq:thermodynamic-state} and the outer-horizon area law give
\begin{equation}
\begin{gathered}
\widetilde M=M,\quad\widetilde J=-J,\quad\widetilde\Omega=-\Omega,\\
\widetilde S=S,\quad\widetilde T=T,\quad
\widetilde M_{\text{irr}}=M_{\text{irr}},\quad
\widetilde E_{\text{rot}}=E_{\text{rot}}.
\end{gathered}
\label{eq:spin-thermodynamic-reflection}
\end{equation}
The mass-generator coefficient $1/b$ is even, so the capture update satisfies
\begin{equation}
\begin{gathered}
\Delta \widetilde{M}=\Delta M=E_1/b,\qquad
\Delta\widetilde{ J}=-\Delta J=-L_1,\\
(\widetilde M',\widetilde J')=(M',-J').
\end{gathered}
\label{eq:spin-update-reflection}
\end{equation}
Since $J$ and $a$ have the same sign, both orientations satisfy $J\Delta J<0$. Whether a finite update crosses zero spin depends on its magnitude.

Among the parameter mapping variables, $j$ reverses sign, whereas $\delta$, $\beta$, and the positive root $x$ remain unchanged. Equation~\eqref{eq:parameter-mapping-cubic} retains the same positive root and branch label, and the reconstruction $a=jM\sqrt{x}/(1+\beta x)$ only reverses the sign of $a$. Reflected initial and final states have equal areas, respectively, so the finite-area condition, $\Delta M_{\text{irr}}$, and $\Delta E_{\text{rot}}$ are identical. Equations~\eqref{eq:eta-eroi}, \eqref{eq:efficiencies}, and~\eqref{eq:run-efficiencies} then give equal single-event efficiencies, cumulative EROI, and EUE.

A repetitive process inherits this result step by step, provided the initial states, particle inputs, geometric branches, outer boundaries, and termination criteria are all reflected consistently. In particular, nonzero $\bar\ell_1$ must reverse sign when $a$ does. Keeping the angular-momentum sign fixed changes the particle's rotation relative to the black hole and defines a different input configuration.

\subsection{Electromagnetic orientation}

The correspondence above holds signed $B$ fixed. For the full electromagnetic field, its orientation must also be tracked. Remove the closed gauge term $\mathbf A_0$ from the potential one-form of the original solution and write $\overline{\mathbf A}=\mathbf A-\mathbf A_0$ and $\mathbf F=\mathrm{d}\overline{\mathbf A}$. For $B\ne0$, the temporal component of the nongauge part is odd in $a$ and the azimuthal component is even, while both are odd in $B$~\cite{ovcharenko_kerr-newman_2026}. Axial reflection therefore gives
\begin{equation}
\begin{gathered}
\mathrm{P}^*\mathbf F(m,-a,B)=-\mathbf F(m,a,B),\\
\mathrm{P}^*\mathbf F(m,-a,-B)=\mathbf F(m,a,B).
\end{gathered}
\label{eq:spin-field-reflection}
\end{equation}
Since the metric contains only $B^2$, the correspondence for the full metric and electromagnetic field is $(a,B,\phi)\mapsto(-a,-B,-\phi)$. At $B=0$, the field vanishes and the Kerr correspondence is recovered. The particles studied here are neutral, so their motion and energy budget depend on the metric, and Eq.~\eqref{eq:spin-reflection-map} at fixed $B$ is sufficient to compare both spin orientations. Extension to charged processes must also account for the electromagnetic field and its coupling to the particles.
\bibliographystyle{apsrev4-2}
\bibliography{references}
\end{document}